\documentclass[11pt,a4paper]{article}
\pdfoutput=1
\usepackage{jheppub}
\usepackage{amsfonts}
\usepackage{bbm}
\usepackage{graphicx}
\usepackage{caption}
\usepackage{subcaption}

\def\v{\varphi}

\def\l{\lambda}    
\def\m{\mu}

\def\r{\rho}                                     
\def\s{\sigma}                                  
\def\t{\tau}

\def\z{\zeta}
\def\vth{\vartheta}

\def\D{\Delta}

\newcommand{\CC}{\mathcal{C}}

\newcommand{\CF}{\mathcal{F}}
\newcommand{\CS}{\mathcal{S}}

\newcommand{\CK}{\mathcal{K}}
\newcommand{\CL}{\mathcal{L}}
\newcommand{\CM}{\mathcal{M}}
\newcommand{\CN}{\mathcal{N}}

\newcommand{\CH}{\mathcal{H}}

\renewcommand{\Im}{{\rm Im}}

\newcommand{\IZ}{\mathbb{Z}}

\newcommand{\ndt}{\noindent}

\def\p{\partial}

\def\bea{\begin{eqnarray}}
\def\eea{\end{eqnarray}}
\def\be{\begin{equation}}
\def\ee{\end{equation}}
\def\ba{\begin{align}}
\def\ea{\end{align}}

\newcommand{\bem}{\begin{pmatrix}}
\newcommand{\eem}{\end{pmatrix}}

\def\={\;  = \;}
\def\+{\, + \,}

\def\wt{\widetilde}
\def\wh{\widehat}
\def\bar{\overline}

\def\rt2{\sqrt{2}}

\renewcommand{\Im}{\mbox{Im}}

\definecolor{orange}{RGB}{255,60,0}

\newcommand{\nv}{n_\text{v}}

\preprint{Nikhef-2015-045}

\title{Single-centered black hole microstate degeneracies from instantons in supergravity}

\author[a]{Sameer Murthy}
\author[b]{and Valentin Reys}
\affiliation[a]{Department of Mathematics, King's College London,
The Strand, London WC2R 2LS, UK}
\affiliation[b]{Nikhef theory group, Science Park 105,
 1098 XG Amsterdam, The Netherlands}

\emailAdd{sameer.murthy at kcl.ac.uk, vreys at nikhef.nl}

\abstract{
We obtain holographic constraints on the microscopic degeneracies of black holes by 
computing the exact macroscopic quantum entropy using localization, 
including the effects of string worldsheet instantons in the supergravity effective action.
For $\frac14$-BPS black holes in type II string theory on~$K3 \times T^{2}$, the constraints can be explicitly checked
against expressions for the microscopic BPS counting functions that are known in terms of certain mock modular forms. 
We find that the effect of including the infinite sum over instantons in the 
holomorphic prepotential of the supergravity 
leads to a sum over Bessel functions with successively sub-leading arguments as in the Rademacher 
expansion of Jacobi forms---but begins to disagree with such a structure near an order where  
the mock modular nature becomes relevant. 
This leads to a systematic method to recover the polar terms of the microscopic degeneracies from 
the degeneracy of instantons (the Gromov-Witten invariants). We check explicitly that our formula agrees 
with the known microscopic answer for the first seven values of the magnetic charge invariant.
}

\begin{document}

\maketitle

\section{Introduction and summary \label{sec:intro}}

Accounting for the thermodynamic entropy of a black hole as the statistical entropy of a microscopic ensemble 
has been one of the important successes of string theory. This idea can be made very precise in theories with supersymmetry 
because we can decouple the near-horizon configuration of a BPS black hole as an independent 
quantum system with~$AdS_{2}$ boundary conditions. In particular, we can define the exact quantum entropy of a black hole
as a functional integral in the gravitational theory with boundary conditions set by the  
attractor mechanism~\cite{Sen:2008vm}.
The exact computation of such functional integrals has been made possible  due to the powerful technique of 
localization~\cite{Pestun:2007rz} applied to supergravity~\cite{Banerjee:2009af, Dabholkar:2010uh, Dabholkar:2011ec}.
Although some important hurdles remain to be crossed in this program, we can---with some explicitly-stated assumptions---begin 
to write formulas for the perturbatively exact quantum entropy of a supersymmetric black hole in string theories with 8 
supercharges in four dimensions~\cite{Murthy:2015yfa}. We can thus compare the microscopic and macroscopic entropy 
formulas at a very precise level, pushing forward the earlier ideas 
of~\cite{LopesCardoso:1998wt,LopesCardoso:2000qm,Ooguri:2004zv,Sen:2005wa,Castro:2008ys,Dabholkar:2005dt,Shih:2005he,Cardoso:2006xz,Denef:2007vg}.

The best understood situation is that of $\frac18$-BPS black holes in maximally supersymmetric ($\CN=8$) theories.
Localization reduces the full perturbative path integral in these theories to a one-dimensional
integral which is simply the integral representation of a modified $I$-Bessel function. 
Going further, one can also identify all non-perturbative saddle-points of the full path integral~\cite{Banerjee:2008ky, Murthy:2009dq}
and compute the contributions of fluctuations around them~\cite{Dabholkar:2014ema}. 
The exact non-perturbative expression for the black hole entropy is thus given by an infinite sum over different saddle-points 
yielding a corresponding infinite sum over $I$-Bessel functions with successively suppressed arguments, which 
add up to precisely the integer degeneracies of the microscopic ensemble computed in~\cite{Maldacena:1999bp}.

This remarkable manner in which continuum gravity arranges integer black hole degeneracies 
relies on the equally remarkable successive approximation of an integer in terms of complex analytic functions---eventually arriving 
at a convergent analytic series. This formula is well-known in analytic number theory as the \emph{Hardy-Ramanujan-Rademacher 
expansion}.
It is a consequence of the modular symmetry of the corresponding microscopic ensemble of the black hole constituents. 
This modular symmetry of the black hole ensemble is, however, special to~$\CN=8$ string theory. 
In theories with lower supersymmetry, there are gravitational configurations other than the black hole that contribute 
to the full entropy formula~\cite{Denef:2000nb, Denef:2007vg} 
(unlike the case for~$\CN=8$ string theories~\cite{Dabholkar:2009dq}), and 
isolating the microstates belonging to the black hole will, in general, destroy modularity. 

We have learnt about many aspects of the modular behavior of the microscopic partition functions 
in the generic setting of~$\CN=2$ theories 
based on the modular nature of the effective strings when black holes descend from wrapped strings, 
and from the spacetime 
duality symmetries of the underlying theory~\cite{Maldacena:1996gb,Dijkgraaf:2000fq,Gaiotto:2006wm,
deBoer:2006vg,Manschot:2007ha,Denef:2007vg,Manschot:2009ia,Alexandrov:2012au}. 
However, the counting function of microstates of a \emph{single} black hole is still not understood in general, and 
in particular, it is not clear to what extent the modular symmetry of the original counting function has any remnant 
in the single-center black holes.
In this paper we begin to address this problem from the point of view of the bulk gravitational theory.

The main point that we make here is that localization allows us to compute the perturbatively exact 
macroscopic formula for the black hole entropy. This formula is a very good analytic approximation to the microscopic 
degeneracies of the single-center black hole, and thus constrains the modular nature of their generating function. 
Under explicit assumptions about the prepotential and the functional integral measure in the language of effective supergravity, 
the exact macroscopic entropy has a structure similar to the 
Rademacher expansion of modular forms. 
As was already derived in~\cite{Dabholkar:2005dt,Denef:2007vg}, following the OSV formula~\cite{Ooguri:2004zv}, 
the leading approximation to the degeneracy is given by a Bessel function with argument equal to a 
quarter of the area of the black hole, in the two-derivative approximation to the Wilsonian effective action of supergravity. 
Here we go beyond the leading order and show that including the infinite series of instanton effects in the 
holomorphic prepotential leads to a finite series of sub-dominant Bessel functions.

We illustrate this formula in the concrete setting of the~$\CN=4$ string theory obtained as a type II 
compactification on~$K3 \times T^{2}$. In this situation we have a complete knowledge of the non-perturbative 
prepotential in the supergravity theory, as well as that of the microscopic BPS counting function for~$\frac14$-BPS 
states (see~\cite{Sen:2007qy}). Further, 
it is known~\cite{Dabholkar:2009dq} that the only configurations, apart from dyonic~$\frac14$-BPS black holes, that contribute to the 
relevant supersymmetric index are two-centered black holes which are each~$\frac12$-BPS. Subtracting this two-centered 
contribution leads, as expected, to a breaking of modular symmetry for the single-centered black hole degeneracies of interest. 
It was shown in~\cite{Dabholkar:2012nd} that this breaking of modular symmetry happens in a very special manner 
and the single-centered black hole degeneracies are coefficients of mock modular 
forms~\cite{Zwegers:2008zna,Zagier:2007}. 
As a consequence, analytic number-theoretic expressions for the degeneracies can be resurrected---at the expense of 
some modifications to the formula due to the mock nature of the partition functions~\cite{Bringmann:2010sd}. 

We find that the macroscopic answer in the~$K3 \times T^{2}$ theory has the following structure.
The prepotential of the theory is exact at one-loop order. The one-loop contribution to the 
prepotential depends only on a special modulus in the theory~$S=-iX^{1}/X^{0}$, and it 
can be expanded as an infinite series in powers of the type~$e^{-nS}$, where~$n$ is identified as the instanton number. 
The zero-instanton sector gives rise to the leading $I$-Bessel-function in the Rademacher expansion of the microscopic 
theory. 
In addition, the contribution from each of the infinite instanton sectors has the right structure to be identified 
with an $I$-Bessel function---seemingly leading to a badly divergent contribution to the answer. 
However, the choice of integration contour ensures that one gets sub-leading $I$-Bessel functions only until a 
certain value of the instanton number, beyond which one obtains exponentially suppressed terms.  

The supergravity partition function can thus be expressed as a sum of Bessel functions with 
successively sub-leading arguments, with exactly the same arguments of the Bessels as those that 
appear in the Rademacher expansion of a Jacobi form. 
Quite remarkably, we find that the coefficients of the Bessel functions also agree exactly for the first 
many Bessel functions---and begin to deviate from the Rademacher expansion of a true Jacobi form 
exactly when we expect them to do so based on the above analysis due to the mock modular nature! 
This shows that the supergravity answer is sensitive to the polar coefficients of the microscopic function \emph{including}
the coefficients of the \emph{mock modular} part. 
This looks to be the beginning of the answer to the question ``How does the continuum supergravity 
know about the mock modular nature of the black hole partition function?''

Going further, we find a compact formula~\eqref{cdrel} for the polar terms of the microscopic black hole 
degeneracies in terms of the degeneracies of the worldsheet instantons. 
As we know quite well by now, the polar terms of a modular form completely control the full function, so our 
formula implies that we can predict all the black hole degeneracies purely from the knowledge of the worldsheet 
instanton degeneracies in the effective action of gravity. 
This idea---that black hole degeneracies can be derived in terms of the instanton degeneracies 
(i.e.~the Gromov-Witten invariants)---was of course one of the main themes of the OSV formula 
and its refinements~\cite{Denef:2007vg}. 
The fact that both the macroscopic and microscopic answers are known and are non-trivial in 
the~$\CN=4$ situation gives us a nice set up to formulate and check these interesting ideas \emph{exactly}. 
We stress that our formula agrees with the microscopic degeneracies of the \emph{single-centered} black hole, 
i.e.~only after we have subtracted the two-centered degeneracies from the full microscopic partition function. 

We also point out a potential interest from a mathematical point of view---namely that our results look like the beginning 
of a consistent large-charge expansion for the coefficients of meromorphic Siegel modular forms which, in contrast with 
the Rademacher expansion for (mock) modular and Jacobi forms, is not really understood in the mathematics literature. 
In order to complete this analysis, we need to classify and consider the effect of all gravitational saddle-points 
with~$AdS_{2}$ boundary conditions (as was done in~\cite{Dabholkar:2014ema} for the~$\CN=8$ theory). 
We leave this interesting problem for the future. 

Our  results thus bring us a step closer in the comparison of the microscopic and macroscopic  
exact degeneracy formulas for~$\CN=4$ string theories. 
It would be interesting to extend this to more general~$\CN=2$ situations. In particular, 
good progress has been made in computing the exact prepotential in F-theory compactifications on 
elliptically fibered Calabi-Yau three-folds~\cite{Huang:2015sta} which indeed has a special modulus
which acts as a modular parameter.  
The problem of disentangling the exact single-centered black hole degeneracy was studied in this context 
in~\cite{Haghighat:2015ega}, but a full answer was not found. 
It would be very interesting if the macroscopic constraints we bring in threw new light on this problem. 

This paper is organized as follows.
In~\S\ref{sec:deg} we present the microscopic counting formula for the quantum degeneracies of 
single-centered BPS black holes in~$\CN=8$ and~$\CN=4$ string theories, in terms of 
Jacobi forms, and mock Jacobi forms, respectively. 
In~\S\ref{locsugra} we present the macroscopic quantum entropy formula in supergravity 
computed using localization methods and show that the leading Bessel function is correctly reproduced 
from the tree-level Wilsonian effective action in the functional integral. 
In~\S\ref{ContourPres} we include the instanton effects into the functional integral, and show that there is 
a finite series of sub-leading Bessel functions which has the same structure as the Rademacher expansion
of Jacobi forms. In~\S\ref{microforms} we perform an exact comparison of this expansion with the 
microscopic degeneracy formula and show how the supergravity theory is sensitive to the mock nature. 
In Appendix~\ref{mockapp} we present some facts about mock Jacobi forms that we use in this paper. 
In Appendix~\ref{app:Iu}  we discuss some details of the contour prescription that we use in~\S\ref{ContourPres}.

\vspace{0.2cm}

\ndt \emph{Note added}: While this paper was being prepared for publication, we received~\cite{Gomes:2015xcf} 
on the arXiv. 
As we use some of the technical analysis presented in that paper, we present a brief review of the relevant parts 
in~\S\ref{ContourPres} and a refinement of some of the details in Appendix~\ref{app:Iu}.
We note that the aims and the main results of the two papers concern
different topics. In particular, we do not aim to derive the measure of the supergravity localization formula 
in the present paper. Our focus here, as sketched above, is to derive an explicit formula relating 
the instanton degeneracies in supergravity and the microscopic black hole degeneracies. 
 
\section{Single-center black hole degeneracy and (mock) Jacobi forms} \label{sec:deg}

In this section we introduce the microscopic degeneracy formulas for the supersymmetric black holes that we 
study in this paper.
We then present some details of the automorphic symmetry properties of the corresponding generating functions, 
which leads to analytic formulas for the degeneracies of a single-centered black hole.
We begin by discussing the well-understood example of~$\frac18$-BPS black holes in~$\CN=8$ string theory 
as an illustration of the ideas. In this case the degeneracies are Fourier coefficients of a classical Jacobi form. 
We then move to the~$\frac14$-BPS black holes in~$\CN=4$ string theory where there are subtleties due to 
wall-crossing and the black hole degeneracies are coefficients of mock Jacobi forms. 
Here we will review the statements relevant to this paper and refer the reader interested in more details of these functions 
to~\cite{Dabholkar:2012nd}.

\subsection*{$\frac18$-BPS black holes in $\CN=8$ string theory}

We begin by considering type II string theory compactified on~$T^{6}$. 
At low energies the effective description of the theory is given by~$\CN=8$ supergravity in four dimensions. 
This theory has a macroscopic~$\frac18$-BPS black hole solution carrying electric and magnetic charges under the various 
gauge fields in the theory. The~$\CN=8$ string theory has an~$E_{7,7}(\IZ)$ duality group with a 
duality invariant~$\D$ that is quartic in the charges, and the entropy of the black hole depends only on~$\D$.
In order to compute the microscopic degeneracies, one goes to a particular duality frame in which there is an
explicit description of the charges of the black hole as charges in the microscopic string theory. A simple description 
consists of at least four charges which can be represented as follows. 
Writing~$T^{6} = T^{4} \times S^{1} \times \wt S^{1}$, one has a D1-brane and a D5-brane wrapped 
on~$S^{1}$ with momentum~$n$ on~$S^{1}$ and one unit of Kaluza-Klein monopole
charge on~$\wt S^{1}$. In addition, one can turn on a fifth charge~$\ell$ which corresponds to the momentum 
around~$\wt S^{1}$. The duality invariant in this configuration is~$\D=4n-\ell^{2}$.

Using this description one can compute the BPS partition function which is the generating 
function of the microscopic index of~$\frac18$-BPS states in the theory 
\be
  Z^\text{BPS}(\t,z) \= \sum_{n, \ell \, \in \, \IZ} c(n, \ell)\,q^n\,\z^\ell \, , 
\ee
which has a simple explicit form in terms of theta and eta functions~\cite{Maldacena:1999bp}:
\be \label{phi21}
Z^\text{BPS}(\t,z) \=  \v_{-2,1}(\t,z) \, := \, \frac{\vartheta_{1}(\t,z)^{2}}{\eta(\t)^{6}} \, . 
\ee
The black hole degeneracies are related to the index of~$\frac18$-BPS states in the 
theory~\cite{Sen:2009vz,Dabholkar:2010rm} as:
\be
d(n,\ell) \= (-1)^{\ell} \, c(n,\ell) \, . 
\ee

\subsection*{Jacobi forms}

The function~$\v=\v_{-2,1}$ is an example of a Jacobi form of weight~$k=-2$ and index~$m=1$ whose defining property is the 
following two transformations. It is ``modular in~$\t$'', i.e.~it transforms under the modular group as
 \be\label{modtransform}  \v\Bigl(\frac{a\t+b}{c\t+d},\frac{z}{c\t+d}\Bigr) \= 
   (c\t+d)^k\,e^{\frac{2\pi imc z^2}{c\t+d}}\,\v(\t,z)  \qquad \forall \quad
   \Bigl(\begin{array}{cc} a&b\\ c&d \end{array} \Bigr) \in SL(2; \IZ) \, ,
\ee
and ``elliptic in~$z$'', i.e.~it transforms under the translations of $z$ by $\mathbb{Z} \tau + \mathbb{Z}$ as
  \be\label{elliptic}  \v(\t, z+\l\tau+\mu)\= e^{-2\pi i m(\l^2 \t + 2 \l z)} \v(\t, z)
  \qquad \forall \quad \l,\,\m \in \IZ \, . \ee

These symmetry properties are very powerful and they give us great control over its Fourier coefficients
\be\label{fourierjacobi}
  \v(\t,z) \= \sum_{n, \ell \, \in \, \IZ} c(n, \ell)\,q^n\,\z^\ell \, . 
\ee
As a simple example, the elliptic transformation property~\eqref{elliptic} implies that the Fourier coefficients 
of a Jacobi form of index~$m$ obey the property 
 \be\label{cnrprop}  c(n, \ell) \= C_{\ell}(4 n m - \ell^2) \ ,
  \quad \mbox{where} \; C_{\ell}(\D) \; \mbox{depends only on} \; \ell \, \text{mod}\, 2m \ . \ee
The coefficients~$C_{\mu}(\D)$ for fixed~$\mu$ are the Fourier coefficients of modular forms. 
Indeed the~$\CN=8$ black hole degneracies~$d(\D)$ can be written in terms of the coefficients~$C_{\ell}(\D)$ 
of the Jacobi form~\eqref{phi21} as:
\be
d(\Delta) =  (-1)^{\Delta +1} C_{\mu} (\Delta) \, , \quad \text{with~$\mu=\D$ mod 2} \, ,
\ee
which is a manifestation of the physical~$U$-duality symmetry. 

The precise mathematical definition of Jacobi forms~\cite{Eichler:1985ja} includes some technical conditions on the growth 
of the Fourier coefficients, in addition to the transformation formulas~\eqref{modtransform}, \eqref{elliptic}. 
There are two types of Jacobi forms that will be relevant to us in this paper. The first is a \emph{weakly holomorphic}
Jacobi form, which means that the Fourier expansion in~\eqref{fourierjacobi} obeys~$n \ge -n_{0}$ for a fixed positive~$n_{0}$. 
This implies that there are only a finite number of terms with Fourier coefficients for negative 
values of~$\D$, and these are called the \emph{polar} terms in the Fourier expansion of the Jacobi form. 
The second type is that of~\emph{weak Jacobi forms} which means that~$n_{0}=0$ above. 
We refer the reader to~\cite{Eichler:1985ja} for a detailed theory of these functions. 

The modular transformation property~\eqref{modtransform} is so powerful that one has an analytic formula 
for all the coefficients of a Jacobi form in terms of its polar coefficients. 
This formula, called the Hardy-Ramanujan-Rademacher formula, takes the form 
of an infinite convergent sum of Bessel functions, and is established by the so-called circle 
method in analytic number theory (see \cite{Dijkgraaf:2000fq} for a nice exposition). 
The formula for the coefficients  $C_{\ell}(\D)$ of a Jacobi form of weight $w+1/2$ and index $m$, with~$\D=4mn-\ell^{2}$, has the following form:
 \be\label{radi}
C_\ell (\D) = (2\pi)^{2-w} \sum_{c=1}^\infty 
  c^{w-2} \sum_{\wt\ell \in \IZ/2m \IZ} \, \sum_{\wt \D < 0} \, 
C_{\wt\ell}(\wt \D) \,  
K\ell(\D,\ell,\wt \D,\wt\ell;c) \, \left| \frac{\wt \D}{4m} \right|^{1-w} \, \wt I_{1-w}
 \Bigl( {\pi\over m c} \sqrt{| \wt \D|  \D}
\Bigr) \, ,
\ee
where 
\begin{equation}\label{intrep}
 \wt{I}_{\rho}(z)=\frac{1}{2\pi
i}\int_{\epsilon-i\infty}^{\epsilon+i\infty} \, \frac{d\s}{\s^{\r +1}}\exp \Bigl( {\s+\frac{z^2}{4\s}} \Bigr) \, ,
\, 
\end{equation}
is called the modified Bessel function of index $\r$, and is 
related to the standard Bessel function of the first kind $I_{\rho}(z)$ by
\be
\wt I_{\rho}(z) = \big(\frac{z}{2} \big)^{-\rho} I_{\rho}(z) \, .
\ee
The coefficients $ K\ell(\D,\ell,\wt \D,\wt\ell;c)$ are the so-called generalized Kloosterman sums~\cite{Dijkgraaf:2000fq},
and they are essentially sums of phases, thus carrying practically no entropy.  For~$c=1$, they are given by:
\be \label{Kloosc1}
K\ell(\D,\ell,\wt \D,\wt\ell;c=1) \= S_{\ell \wt\ell}^{-1} \= \sqrt{\frac{2}{m}}\,e^{i\pi(m-\ell')\frac{\ell}{m}} \, . 
\ee
The remarkable thing about Formula~\eqref{radi} is that the coefficients~$C_{\ell}(\D)$ for~$\D>0$ are completely determined by      
the coefficients~$C_{\wt\ell}(\wt \D)$ associated to the so-called polar terms~$q^{\wt \D}$ with~$\wt \D<0$, which are finite 
in number. The asymptotic formula of the Bessel function~$I(z) \sim e^{z}$ for large~$z$ shows that the terms with~$c>1$
are exponentially suppressed compared to the leading~$c=1$ terms.

The~$\CN=8$ black hole partition function~$\v_{-2,1}$~\eqref{phi21} is a weak Jacobi form.
There is only one polar coefficient with~$\wt \D = -1$, and so the Rademacher formula simplifies to:
\be\label{rademsp} 
 C(\D) =   2{\pi} \, \big( \frac{\pi}{2} \big)^{7/2} \, \sum_{c=1}^\infty 
  c^{-9/2} \, K_{c}(\D) \; \wt I_{7/2} \big(\frac{\pi \sqrt{\D}}{c} \big)  \, .
\ee
Here~$K_{c}$ is a particular combination of the Kloosterman sums with the property~$K_{1}(\D)=1$. 
As was shown in~\cite{Dabholkar:2011ec}, the leading $c=1$ Bessel function can be recovered as 
the exact functional integral of supergravity with~$AdS_{2} \times S^{2}$ boundary conditions. 
It can thus be interpreted as the all-order perturbation theory result for the quantum entropy, including all 
perturbative quantum corrections to the Bekenstein-Hawking formula. 
The exponentially sub-leading terms with~$c>1$, including the Kloosterman sums, can also be recovered as exact functional 
integrals over different orbifold sectors of~$AdS_{2} \times S^{2}$~\cite{Dabholkar:2014ema}.

\subsection*{$\frac14$-BPS black holes in $\CN=4$ string theory}

Now we come to the~$\CN=4$ theory. 
Consider type II string theory compactified on~$K3 \times T^{2}$ or, equivalently, heterotic string theory on~$T^{6}$. 
At low energies the effective description of the theory is given by~$\CN=4$ supergravity coupled to 28~$\CN=4$ 
gauge field multiplets specified by the compactification.
The quarter-BPS black holes carry electric and magnetic charges~$(Q_{e}^{i}, Q_{m}^{i})$ ($i=1,\cdots, 28$), 
under these gauge fields, where~$i$ is a vector index under the T-duality 
group~$SO(6,22)$, and~$(Q_{e},Q_{m})$ transform as a doublet under the S-duality group~$SL(2,\IZ)$. 
The U-duality group of the theory is~$SL(2,\IZ) \times SO(6,22)$  

One-fourth BPS dyonic states in the theory are completely labelled by the three continuous 
T-duality invariants:
\be \label{nlm}
(Q_{e}^{2}/2, Q_{e} \cdot Q_{m}, Q_{m}^{2}/2) \, \equiv \, (n,\ell,m) \, ,
\ee
and, in addition, some discrete charge invariants~\cite{Banerjee:2008ri}. 
As for the~$\CN=8$ example we write the compactification manifold as $K3 \times S^{1} \times \wt S^{1}$,
and we can choose a duality frame in which the black hole consists of
the D1-D5-P system wrapping~$K3 \times S^{1}$ with~$Q_1$ D1-branes,~$Q_5$ D5-branes, one unit of KK-monopole charge and~$\ell$ units 
of momentum on~$\wt S^{1}$. The charge invariants are~($Q_e^2/2 = n$, $Q_e\cdot Q_m=\ell$, $Q_m^2/2 = Q_1 Q_{5}$). 
The exact microscopic counting formula for the index of a generic one-fourth BPS state has been worked out 
completely~\cite{Dijkgraaf:1996it,Shih:2005uc, David:2006yn, Banerjee:2008pu, Dabholkar:2008zy}. 
For charges where the discrete invariants are trivial, the BPS indexed partition function is given by
\be
Z^\text{BPS}(\t,z,\s) \= \frac{1}{\Phi_{10}(\t,z,\s)} \, ,
\ee
where we now have three chemical potentials that couple to the three T-duality invariants. 
The function~$\Phi_{10}$ is the Igusa cusp form, which is the unique Siegel modular form of weight~10. 
The microscopic degeneracy is given by the so-called DVV formula~\cite{Dijkgraaf:1996it}:
\be
\label{eq:DVV}
d(n,\ell,m) = (-1)^{\ell + 1}\int_\mathcal{C} d\t dz d\s \, \frac{e^{-i\pi\left(\tau n + 2z \ell + \sigma m \right)}}{\Phi_{10}(\t,z,\s)} \, ,
\ee
with a contour~$\CC$ that was spelled out in~\cite{Cheng:2007ch}.

\subsection*{Mock Jacobi forms}

There is an important new physical phenomenon that arises in the~$\CN=4$ theory as compared to the~$\CN=8$ theory. 
While the microscopic index that counts~one-eighth-BPS
states  in the~$\CN=8$ theory only gets contributions from single-centered black holes, the corresponding index that 
counts~quarter-BPS states in the~$\CN=4$ theory gets contributions from single-centered black holes as well as 
two-centered black hole configurations, depending on the value of the moduli at infinity~\cite{Dabholkar:2009dq}. 
This ambiguity is captured in the DVV formula by the choice of contour in~\eqref{eq:DVV}, 
which depends on the moduli fields at infinity~\cite{Dabholkar:2007vk, Cheng:2007ch}. 
Choosing the moduli to be at the attractor point yields the pure single-centered black hole degeneracies. 
Doing so, however, destroys the modular symmetry. From a physical point of view this breaking is related to the fact that 
we are throwing away a part of the spectrum of the theory. From a mathematical point of view it is because the 
partition function~$1/\Phi_{10}$ is a \emph{meromorphic} function with poles in the bulk of the Siegel upper half plane. 

Without the powerful handle given by the modular symmetry, it looks at first sight like the program followed 
to interpret the microscopic degeneracies in supergravity will not work. 
In particular, we do not know how to write down an analytic expansion like~\eqref{rademsp} for 
the~$\CN=8$ black hole case.
This problem was solved in~\cite{Dabholkar:2012nd} as we now briefly summarize. (We give more details in 
Appendix~\ref{mockapp}.) We can perform one of 
the three Fourier expansions in~\eqref{eq:DVV} near~$\s\to i\infty$ to obtain:
\be\label{reciproigusa}
  \frac 1{\Phi_{10}(\t, z, \sigma)} \= \sum_{m\geq -1} \psi_m (\t,z) \, e^{2 \pi i m \s}  \, .
\ee
The functions~$\psi_{m}$ are Jacobi forms of weight~$-10$ and index~$m$ that are meromorphic (in~$z$). 
These contain the degeneracies of states with magnetic charge~$m$, including both single and two-centered 
black holes. 
The single-centered black hole degeneracies are found by subtracting the generating function of two-centered degeneracies 
(called $\psi_{m}^\text{P}$) from~$\psi_{m}$. The difference, called the finite or Fourier part of $\psi_{m}$
\be \nonumber
\psi_{m}^\text{F} = \psi_{m} - \psi_{m}^\text{P} \, ,
\ee
is holomorphic in $z$, and has an unambiguous Fourier expansion:
\be \label{psiF}
\psi_{m}^\text{F}(\t,z) \= \sum_{n,\ell} \, c^\text{F}_{m}(n,\ell) \, q^{n} \, \zeta^{\ell} \, .
\ee
It was shown in~\cite{Dabholkar:2012nd} that: 
\begin{enumerate}
\item[(i)] The microscopic indexed degeneracies~$d(n,\ell,m)$ of the single-centered black holes (i.e.~corresponding to  
the attractor contour) are precisely related to the Fourier coefficients of this function 
\be 
d(n,\ell,m) = (-1)^{\ell+1} c^\text{F}_{m}(n,\ell) \, , 
\ee
\item[(ii)] The function~$\psi^\text{F}_{m}(\t,z)$ is a \emph{mock Jacobi form}.
\end{enumerate}

The meaning of the word \emph{mock} is that 
the transformation rule~\eqref{modtransform} is modified. The functions~$\psi^\text{F}_{m}$ themselves are not modular,
but one can add a correction term called the \emph{shadow} to get completed functions~$\wh{\psi^\text{F}_{m}}$ that 
\emph{are} modular, i.e.~they transform exactly with the rule~\eqref{modtransform}. 
The shadow is a non-holomorphic 
function\footnote{See~\cite{Troost:2010ud,Eguchi:2010cb,Ashok:2011cy,Murthy:2013mya,Ashok:2013pya,Harvey:2014nha,Pioline:2015wza} 
for the physical origin of such non-holomorphic terms from the point of view of conformal field theory. 
Understanding the physical basis of the non-holomorphicity of the specific functions~$\wh{\psi^\text{F}_{m}}$ 
is an interesting open problem.} and leads to a holomorphic anomaly equation as in~\eqref{ddtbarhphi}.
This resurrection of modular symmetry means, in particular, that we can again use the circle method to get a 
formula for the Fourier coefficients. 
This formula differs from that of the analogous formula for true Jacobi forms (the Rademacher expansion)
due to the effect of the shadow term (see \cite{Bringmann:2010sd, Bringmann:2012zr}). 
In order to make sharp estimates about how the asymptotic expansion of mock Jacobi forms differs from 
that of true Jacobi forms, we need to know the explicit expressions of the mock Jacobi forms in question. 
This is a fairly complicated question but it has been addressed and solved in~(\cite{Dabholkar:2012nd}, Chapters 9, 10). 
We provide some relevant details in Appendix~\ref{mockapp}, and here we illustrate the main points with some examples.

In order to present the results, we need to introduce two Jacobi forms 
\be\label{phi2}
A(\tau, z)=\v_{-2,1}(\tau, z) :=  \frac{\vth_1^2(\t, z)}{\eta^6(\t)} \, ,
\ee
 \be\label{phi0}
 B(\tau, z)=\v_{0, 1} (\t, z) := 4 \left( \frac{\vth_2^2(\t, z)}{\vth_2^2(\t)} +
    \frac{\vth_3^2(\t, z)}{\vth_3^2(\t)} +\frac{\vth_4^2(\t, z)}{\vth_4^2(\t)} \right) \, ,
\ee
where $\vartheta_{i}, i=1,\dots,4$ are the four classical Jacobi theta functions, with coefficients that are monomials in 
the Eisenstein series~$E_{4}$, $E_{6}$. These two Jacobi forms generate the ring of 
all weak Jacobi forms of even weight over the ring of modular forms~\cite{Eichler:1985ja}. 
The word ``weak'' here refers to a growth condition on the functions, and it means in particular that for 
large values of~$\D=4mn-\ell^{2}$, the coefficients grow as (see Appendix~\ref{mockapp}) 
\be \label{Jacgrowth}
c(n,\ell) \simeq \exp(\pi\sqrt{4mn-\ell^{2}}) \, . 
\ee

The functions~$\psi^\text{F}_{m}$ can be worked out explicitly (see~\cite{Bringmann:2012zr}) for a given value of~$m$. 
The first couple of cases are:
\bea \label{psiF12}
\psi_1^\text{F} &= & \frac1{\eta(\t)^{24}} (3 E_4A- 648\mathcal{H}_{1}) \, , \label{m1}\\
 \psi_2^\text{F}&= & \frac{1}{3\eta(\t)^{24}}\big(22E_4 AB-10 E_6A^2-9600 \mathcal{H}_{2} \big) \, .\label{m2}
\eea
Here the functions~$\CH_{1}$, $\CH_{2}$ are mock Jacobi forms whose coefficients 
are linear combinations of the so-called Hurwitz-Kronecker class numbers, whose Fourier coefficients have 
purely polynomial growth. This is representative of the general structure proved in~\cite{Dabholkar:2012nd}:
the mock Jacobi forms~$\psi^\text{F}$ can always be written as a sum of two pieces:~$\v^\text{true}_{2,m}(\t,z)/\eta(\t)^{24}$ 
and~$\v^\text{opt}_{2,m}(\t,z)/\eta(\t)^{24}$. The function~$\v^\text{true}_{2,m}(\t,z)$ is a true weak Jacobi form 
(in particular, we can apply the usual Rademacher expansion~\eqref{radi} to it), and the second is
a mock Jacobi form of a very special kind in that its Fourier coefficients grow extremely slowly. 
In the two examples above, this growth is purely polynomial---that is the case whenever~$m$ is a prime power.
In general, the growth of the coefficients of~$\v^\text{opt}_{2,m}(\t,z)$ goes as
\be \label{optgrowth0}
c^\text{opt}(n,\ell) \sim \exp\bigl(\frac{\pi}{m}\sqrt{4mn-\ell^{2}} \bigr)  \, .
\ee
which can be contrasted with~\eqref{Jacgrowth}. 
What we need is to estimate the growth of the ratios like~$\v^\text{opt}_{2,m}(\t,z)/\eta(\t)^{24}$ that enter our expressions.
Such functions are called \emph{mixed} mock Jacobi forms, and their Rademacher expansion already differs at 
leading order in the asymptotic expansion compared to a true Jacobi form of the same weight and index 
(see Comment 1 below Theorem (1.3) of~\cite{Bringmann:2010sd}). 

We are now ready to reap the benefits of this technical analysis. 
If we want to analyze the Rademacher expansion of the black hole degeneracies encoded in~$\psi_{m}^\text{F}$, we can use 
the usual Rademacher expansion~\eqref{radi} of Jacobi forms \emph{as long as} the growth of Bessel functions in~\eqref{radi}
are larger than the growth of the mixed mock Jacobi forms~$\v^\text{opt}_{2,m}(\t,z)/\eta(\t)^{24}$. 
From what we said above, it is clear that we always have the contribution of the (denoting polynomial prefactors by~$p_{i}$ for 
now)
\be \label{LeadBes}
\text{Leading Bessel:} \qquad p_{0} \, \wt I_{23/2}  \biggl( {2\pi} \sqrt{ (m+4) \Bigl( n -\frac{\ell^{2}}{4m} \Bigr) } \, \biggr) \, ,
\ee
where~$p_{0}=(m+2)\frac{4\pi}{\sqrt{m} }  \bigl(\frac{m+4}{n - \frac{\ell^2}{4m}}\bigr)^{23/4}$
as for a true Jacobi form for any~$m$. This is then followed by the sub-leading Bessel functions in 
the~$c=1$ series of~\eqref{radi}:
\bea
&& \text{Sub-leading $c=1$ series:} \quad
p_{1}  \, \wt I_{23/2}  \biggl( {2\pi} \sqrt{ (\frac{(m-1)^{2}}{m}+4) \Bigl( n -\frac{\ell^{2}}{4m} \Bigr) } \, \biggr) \, + 
\qquad \qquad \qquad  \qquad     \\
&& \qquad \qquad \qquad \qquad \qquad \qquad  \qquad \qquad
\qquad p_{2}  \, \wt I_{23/2}  \biggl( {2\pi} \sqrt{ (\frac{(m-2)^{2}}{m}+4) \Bigl( n -\frac{\ell^{2}}{4m} \Bigr) } \, \biggr) + \cdots \, .
\nonumber
\eea
But we should stop trusting this series when one of two things happen: 
firstly the~$c=2$ term begins to contribute at the order 
\be \label{c2Bes} 
\text{$c=2$ series:} \qquad \wt I_{23/2}  \biggl( {2\pi} \sqrt{ \frac{(m+4)}{4} \Bigl( n -\frac{\ell^{2}}{4m} \Bigr) } \, \biggr) \, .
\ee
Secondly the mock modular terms begin to contribute according to the discussion above.
We need to use a modified Rademacher expansion for the mixed mock Jacobi forms 
as in~\cite{Bringmann:2010sd}. Working out the details of the latter is an interesting problem in analytic number theory 
which we leave for the future (and for the experts!).  

We will use this analysis in \S\ref{microforms} to 
work out some details of when exactly the signature of the mock nature appears in the Rademacher expansion 
on a case-by-case basis for the first few values of~$m$. We now change track and move on to a supergravity 
analysis of the single-center black hole partition function.

\section{Localization in supergravity \label{locsugra}}

We begin this section with a review of the exact computation of the quantum entropy of BPS black holes in 
four-dimensional~$\CN=2$ supergravity following~\cite{Dabholkar:2010uh, Dabholkar:2011ec}.  
We consider the particular case of~$\frac14$-BPS black holes in~$\CN=4$ string theory coming from the 
compactification of Type II string theory on~$K3 \times T^{2}$. 
In the two-derivative limit of supergravity, we show how the functional integral in the 
near-horizon~$AdS_{2}$ reduces to a single Bessel function. 
We then set the stage for the inclusion of instantons in the holomorphic prepotential of the supergravity,
which we will treat in the next section.

The theory under consideration is described by~$\mathcal{N}=2$ superconformal gravity in 4 dimensions 
with the Weyl multiplet coupled to~$\nv +1$ vector multiplets. 
In this theory we consider a BPS black hole solution carrying electric and magnetic charges~$q_{I}$, $p^{I}$. 
In order to compute the exact quantum entropy of this black hole, we use the 
supersymmetric localization technique applied to this problem~\cite{Dabholkar:2010uh, Dabholkar:2011ec}. 
The near-horizon configuration of the classical black hole is found by the attractor mechanism to 
be the~$AdS_{2} \times S^{2}$ geometry with constant electric and magnetic fields, and constant scalar fields: 
\bea \label{attrvalues}
&&ds^2 = v\bigl( d\eta^2 + \sinh^{2} \eta \, d\tau^2 + d\psi^2 + \textnormal{sin}^2 \psi\,d\phi^2 \bigr) \, , \nonumber \\
&&F^I_{\eta\tau} = i e^I_{*} \sinh \eta  \, , \quad F^I_{\psi\phi} = p^I\textnormal{sin}\psi \, , \quad X^{I} = X^{I}_{*} \, .
\eea
Here $X^I_* = \frac{1}{2}(e_*^I + ip^I)$ is the attractor value of the scalar field with $e_*^I$ and $p^I$ the 
electric field and the magnetic charge of the black hole, respectively. 
The electric field is related to the electric charge~$q_{I}$ of the black hole by a Legendre transform~$q_{I}=\p \CL/\p e^{I}_{*}$. 

In order to compute the exact functional integral using localization, we first find the localization manifold~$\CM_{Q}$
which is the locus of all solutions to the BPS equations. 
The results of~\cite{Dabholkar:2010uh, Gupta:2012cy}  
show that only the scalar and auxiliary fields of the vector multiplets are excited on this localization manifold:
\be
X^I = X^I_* + \frac{C^I}{\cosh \eta} \qquad \bar{X}^I = \bar{X}^I_* + \frac{C^I}{\cosh \eta} \qquad 
Y_1^{I\;1} = -Y_2^{I\;2} = \frac{2C^I}{\cosh^2 \eta}\, ,
\ee
while all other fields stay fixed to their classical attractor values~\eqref{attrvalues}.
It is convenient to label the localizing manifold by the variables:
\be
\phi^I \equiv e_*^I + 2C^I \, . 
\ee

The exact quantum entropy of the black hole, as derived in~\cite{Dabholkar:2010uh} is:
\be \label{exactS}
\widehat{W}(q, p) = \int_{\mathcal{M}_{Q}}\exp\big(\mathcal{S}_\text{ren}(\phi, q, p) \big) \, [d \mu(\phi)] \, .
\ee
Here $[d \mu(\phi)]$ is the induced measure from the supergravity, including the 
classical induced measure longitudinal to~$\CM_{Q}$, and a one-loop determinant coming from integration 
over non-BPS directions orthogonal to~$\CM_{Q}$. We will comment on this in the following. 
The function~$\CS_\text{ren}$ is the renormalized action of the theory coming from evaluating the 
full supergravity action on the localizing solutions and following a regularization procedure~\cite{Sen:2008vm, Dabholkar:2010uh} 
to remove the divergences coming from the infinite volume of~$AdS_{2}$. 
  
The full action of~$\CN=2$ supergravity can be divided into two class---terms coming from chiral superspace integrals 
of the holomorphic prepotential, and those coming from full superspace integrals. It was shown in~\cite{Murthy:2013xpa} that 
only the former type of terms contribute to the quantum entropy. These terms are encoded in one holomorphic 
function~$F(X^{I}, \wh{A})$ describing the coupling of the vector multiplets to the background Weyl 
multiplet. 
Here,~$X^I$ is the lowest component of the vector multiplet and~$\wh{A} \equiv (T_{\mu\nu}^-)^2$ is the lowest component 
of the chiral multiplet built as the square of the Weyl multiplet. 
The renormalized action on the localization manifold evaluated for the action governed by~$F$ 
takes the simple form:
\be \label{Sren}
\mathcal{S}_{ren}(\phi, q, p) =  - \pi  q_I   \phi^I + 4 \pi \, \Im  F\Big(\frac{\phi^I+ip^I}{2} \Big) \, . 
\ee
In this equation (and sometimes in the following), we have used the fact that the attractor equations set~$\wh{A}=-64$. 
The prepotential function~$F(X^{I},\wh{A})$ is a homogeneous function of weight 2 
under the scalings~$X^{I} \to \lambda X^{I}$, $\wh{A} \to \lambda^{2} X^{I}$, and it can be expanded as:
\be \label{FAexp}
F(X^{I},\wh{A}) \= \sum_{g=0}^{\infty} \, F^{(g)}(X^{I}) \, \wh{A}^{g} \, .
\ee
The function~$F^{(0)}(X^{I})$ controls the two-derivative interactions, 
and the coefficients~$F^{(g)}$, $g \ge 1$, describe higher derivative couplings in the theory.

Now we consider the~$K3 \times T^{2}$ compactification of the type II theory. Writing this theory as an~$\CN=2$
supergravity yields a field content, in addition to the Weyl multiplet, of vector multiplets, hyper multiplets and 
gravitino multiplets. 
Following the ideas of~\cite{Shih:2005he} one can truncate this theory to an~$\CN=2$ supergravity with a Weyl 
multiplet and~$\nv=23$ vector multiplets. 
In this case the perturbative prepotential has the form:
\be \label{Ftree}
F^\text{tree}(X) = -\frac{X^1}{X^0} \, X^a C_{ab} X^b + \frac{X^1}{X^0} \, ,
\ee
where~$C_{ab}$ is the intersection matrix of~$K3$. In the full theory, this is modified due to the effects of 
worldsheet instantons, as we shall consider in the following.

Within this set up one can solve the exact functional integral explicitly, as we now briefly recall. 
It was argued in~\cite{Dabholkar:2010uh}, based on the structure of the classical metric of the moduli space, 
that the induced measure on the localizing manifold in the large-charge limit is:
\be \label{msr}
[d\mu(\phi)] \= P_{1} \, \frac{1}{p^1\phi^0}\; \prod_{I} d \phi^{I} \, , 
\ee
where the prefactor~$P_{1}$ is a function only of the charges and independent of the coordinates~$\phi^{I}$. 
One generically expects the measure factor to change when we go beyond the tree-level approximation. 
We shall discuss this in the next section.

Under these assumptions, the quantum entropy~\eqref{exactS} takes the form 
\be
\wh{W}^\text{tree}(p,q) = P_{1} \int \frac{d\phi^0 \, d\phi^1}{\phi^0 p^1}  \, \exp\bigl(-\pi\phi^0q_0 \bigr) \, \int  \prod_{a=2}^{\nv}d \phi^{a} \; 
\exp\biggl(-\pi\phi^2q_2 + 4 \pi \, \text{Im} F^\text{tree}\Bigl(\frac{\phi^I+ip^I}{2}\Bigr)\biggr)  \, .  
\ee
From~\eqref{Ftree}, we see that the last $(\nv-1)$ integrals are Gaussian integrals which yield:
\bea
&& \displaystyle{\int}  \prod_{a=2}^{\nv}d \phi^{a} \; \exp\biggl(-\pi\phi^2q_2 + 4 \pi \, \text{Im} F^\text{tree}\Bigl(\frac{\phi^I+ip^I}{2}\Bigr)\biggr) \\
&& \qquad\qquad\qquad
= \left(\frac{\phi^0}{p^1}\right)^{(\nv-1)/2}\! \exp\biggl({\pi\frac{\phi^1}{p^1}p^1 q_2}\biggr) \, 
\exp\biggl({\pi\frac{\phi^1}{\phi^0} \left(\frac{\phi^1}{p^1}+\frac{p^1}{\phi^1}\right)p^aC_{ab}p^b+4\pi\frac{p^1}{\phi^0}}\biggr) \, . 
\nonumber 
\eea
The change of variables
$\tau_1 = \phi^1/\phi^0$, $\tau_2 =  p^1/\phi^0$
yields
\be \label{Wtree}
\wh{W}^\text{tree}(p,q) = P_{1}\displaystyle{\int} \frac{d\t_{1} d\t_{2}}{\tau_2^{(\nv+3)/2}} \,
\exp\Bigl(\frac{\pi}{\tau_2}\left(-p^1q_0+p^1q_2\tau_1+p^aC_{ab}p^b\tau_1^2+(p^aC_{ab}p^b+4)\tau_2^2\right) \Bigr) \, . 
\ee
 
Upon identifying the four-dimensional electric and magnetic charge invariants as
\be
Q_{e}^{2}/2 \equiv -q_0p^1 \, , \qquad  Q_m^{2}/2 \equiv  p^aC_{ab}p^b \, , \qquad  
Q_{e} \cdot Q_{m} \equiv -q_{2} p^{1} \, , 
\ee
and with the 
identification~$(Q_{e}^{2}/2, Q_{e} \cdot Q_{m}, Q_{m}^{2}/2) = (n,\ell,m)$ as in Equation~\eqref{nlm},
this takes the form,
\be  
\wh{W}^\text{tree} (n,\ell,m) \= P_{1} \displaystyle{\int} \frac{d^2\tau}{\tau_2^{(\nv+3)/2}} 
\; \exp\Bigl(\frac{\pi}{\tau_2}\left(n -\ell\tau_1 + m\tau_1^2+(m+4)\tau_2^2\right) \Bigr) \, .
\ee
The $\tau_1$ integral is Gaussian and can be evaluated in a straightforward manner. 
The remaining integral over~$\t_{2}$ can be evaluated
using the contour integral representation of the Bessel function~\eqref{intrep}, 
\be \label{Whatfin}
\wh{W}^\text{tree}(n,\ell,m) \= P_{1} \frac{2\pi}{\sqrt{m} }  \biggl(\frac{m+4}{n - \frac{\ell^2}{4m}}\biggr)^{23/4} 
I_{23/2}\left(2\pi\sqrt{(m+4) \biggl(n - \frac{\ell^{2}}{4m}\biggr)}\right)\, .
\ee
It has been argued recently in~\cite{Gomes:2015xcf} that the prefactor $P_{1}=2m+4$. 
The function~\eqref{Whatfin} then agrees precisely with the leading Bessel function in the Rademacher expansion 
of the microscopic theory~\eqref{radi} with the right weight, argument, and prefactor.

Now we move to the instanton contributions. 
We note that we kept only the perturbative prepotential to first sub-leading order while in general we 
have instanton sums that generate an infinite series of corrections to the prepotential~\eqref{Ftree}.
In general the instantons contribute to all the couplings~$F^{(g)}$. 
In the type II theory on~$K3 \times T^{2}$ the holomorphic prepotential is one-loop exact:
\be \label{Finst}
F(X) \= -\frac{X^1 X^a C_{ab} X^b}{X^0} +  \frac{1}{2\pi i} \CF^{(1)}_{K3\times T^{2}} (X^{1}/X^{0}) \, .
\ee
The one-loop contribution to the prepotential is:
\be \label{F1K3}
\CF^{(1)}_{K3\times T^{2}} (X^{1}/X^{0}) \=  \log\Bigl(\eta^{24} \bigl(X^{1}/X^{0}\bigr) \Bigr) \, ,
\ee
and has the expansion 
\be 
\CF^{(1)}_{K3\times T^{2}} (X^{1}/X^{0}) \= 2\pi i \, \frac{X^1}{X^0}  + \wt \CF^\text{inst} (X^{1}/X^{0}) \, .
\ee
Here the function~$\wt \CF^\text{inst}$ encodes the contributions of worldsheet instantons in the type II theory 
to the prepotential:
\be \label{Ftlinst}
\wt \CF^\text{inst} (\t) \= -\log \prod_{n=1}^{\infty} (1-e^{2\pi i n \t})^{-24} \, .
\ee

A natural question is how to properly take these corrections coming from the instantons into account. 
The instantons can affect the exact answer~\eqref{exactS} in two ways---by the explicit change of the prepotential in the 
exponent of~\eqref{Sren}, and by an implicit effect on the measure of the integral (which was also computed above in the 
zero-instanton sector). This is what we turn to in the following section.

\section{Including instantons in the functional integral \label{ContourPres}}

In this section, we work out the corrections to~\eqref{Whatfin} due to instantons. 
We write:
\be \label{WhatInst}
\wh{W} (n,\ell,m) =  \int_{\gamma} \frac{d^{2} \tau}{\tau_2^{(\nv+3)/2}} \; 
e^{\frac{\pi}{\tau_2}\left(n-\ell\tau_1+m\tau_1^2+m\tau_2^2\right)} \;
M(\t, \overline{\t}) \, e^{-\CF^{(1)}(\tau) - \CF^{(1)}(-\bar{\tau})} \, .
\ee 
Here we have taken into account the explicit effect on the prepotential function:
\be \label{Ffull}
F(X) \= -\frac{X^1 X^a C_{ab} X^b}{X^0} + \frac{1}{2i\pi}\CF^{(1)}\Bigl(\frac{X^1}{X^0}\Bigr) \, ,
\ee
with~$\CF^{(1)}$ given in~\eqref{F1K3} being the one-loop effect (which is exact in this case), 
which contains contributions from an infinite set of worldsheet instantons wrapping the torus. 
Naively the inclusion of all these instantons leads to an infinite series of~$I$-Bessel functions. 
In this section we show that with an appropriate choice of contour~$\gamma$ in~\eqref{WhatInst}
most of these are in fact exponentially suppressed, leading to a \emph{finite} number of Bessel 
functions that contribute to the quantum entropy. This finite sum has precisely the same 
structure as the leading~$c=1$ term in the expansion~\eqref{radi} for Jacobi forms.

We preface the calculation in this section with some remarks on the measure in Equation~\eqref{WhatInst}. 
We have parametrized the effect of instanton corrections on the measure of the functional integral 
by the function~$M(\t, \bar \t)$. 
In~\S\ref{locsugra} we did not take into account the full quantum effects on the measure in the localization computation. 
Indeed one needs to compute the one-loop determinant of the off-shell fluctuations of the non-BPS modes,
and compute the induced measure from the supergravity field space. The former was computed 
explicitly in~\cite{Murthy:2015yfa, Gupta:2015gga} for matter multiplets in~$\CN=2$ supergravity. Further 
it was argued that the symmetries fix the form of the graviton multiplet determinant up to a constant that can be 
matched to an on-shell computation~\cite{Sen:2011ba}.
The latter has been addressed in various papers~\cite{Cardoso:2008fr, LopesCardoso:2006bg, Gomes:2015xcf} although we think it is fair to say that a 
full satisfactory first-principles derivation of this measure has not been reached yet. We do not attempt to solve this problem 
in the current paper. Instead we will use the fact that 
one knows the exact measure factor based on a saddle-point approximation of the DVV 
formula~\cite{Shih:2005he, David:2006yn, LopesCardoso:2004xf}, as we shall now present. 
Note that this is a different expansion compared to~\cite{Dabholkar:2012nd} that is used to compute
the exact single-centered degeneracies. 

We begin with the DVV formula~\eqref{eq:DVV} which is a three-dimensional contour integral\footnote{For the 
next few lines we will use the variables~$(\s,v,\rho)$ instead of~$(\t,z,\s)$ to avoid confusion.}:
\be
d(n,\ell,m) = (-1)^{\ell + 1}\int_{\mathcal{C}} d\s dv d\r \, \frac{e^{-i\pi\left(\s n + 2v \ell + \rho m \right)}}{\Phi_{10}(\r,v,\s)} \, .
\ee
We can perform an exact contour integral in the~$v$-variable which reduces to 
picking up residues at the divisors of $1/\Phi_{10}$ in the Siegel upper-half plane, leaving a two-dimensional integral 
over~$\s,\r$ which are reexpressed as~$\s=\t_{1}+i\t_{2}$, $\r=-\t_{1}+i\t_{2}$. 
The result is~\cite{David:2006yn}:   
\be \label{dmicint}
d(n,\ell,m) \simeq \int_\gamma \frac{d\t_{1} d\t_{2}}{\tau_2^2}\;e^{-F(\tau_1,\tau_2)} \, ,
\ee
where $\simeq$ implies equality up to exponentially suppressed contributions coming from additional poles, 
which we shall suppress from now on. 
The function~$F(\tau_1,\tau_2)$ is given by:
\bea
F(\tau_1,\tau_2) = &-&\frac{\pi}{\tau_2}\left(n-\ell\t_{1}+m(\tau_1^2 + \tau_2^2)\right) + \ln \eta^{24}(\tau_1 + i\tau_2) + \ln \eta^{24}(-\tau_1+i\tau_2) + 12\ln(2\tau_2) \nonumber \\
&-&\ln\left[\frac{1}{4\pi}\left\{26+\frac{2\pi}{\tau_2}(n-\ell \t_{1}+m(\tau_1^2 + \tau_2^2))\right\}\right] \, ,
\eea
and the contour of integration $\gamma$ is required to pass through the saddle-point of $F(\tau_1,\tau_2)$. 
We rewrite this formula by adding the following total derivative\footnote{This trick was independently noted and 
used in~\cite{Gomes:2015xcf}.} 
to the integrand of~\eqref{dmicint},
\be
\label{eq:totdiv}
\displaystyle{\frac{d}{d\tau_2}}\left(\frac{1}{\tau_2^{13}}e^{\frac{\pi}{\tau_2}(n-\ell\t_{1}+m\tau_1^2+m\tau_2^2) - 
\ln\eta^{24}(\t_{1}+i\t_{2}) - \ln\eta^{24}(-\t_{1}+i\t_{2})}\right) \, , 
\ee
which yields (with~$\t=\t_{1}+i\t_{2}$):
\be
\label{eq:davidsen}
d(n,\ell,m) = \frac{1}{2^{12}}\int_\gamma \frac{d^2\tau}{\tau_2^{13}}\left(m+E_2(\tau)+E_2(-\bar{\tau})\right)(\eta^{24}(\tau)\eta^{24}(-\bar{\tau}))^{-1}e^{\frac{\pi}{\tau_2}(n - \ell\tau_1+m\tau_1^2+m\tau_2^2)} \, , 
\ee
where $E_2$ is the Eisenstein series of weight 2 related to the Dedekind eta function as:
\be \label{E2etarel}
E_2(\t) \= \frac{1}{2\pi i} \,\frac{d}{d\tau}\log\eta^{24}(\t) \, .
\ee
Comparing this to our parametrization~\eqref{WhatInst}, we obtain:
\be \label{ME2}
M(\t,\overline{\t}) \= \frac1{2^{12}}(m + E_2(\t) + E_2(-\overline{\t})) \, . 
\ee

We note that~$M$ can be written in terms of the generalized K\"{a}hler potential defined as
\be \label{defCK}
e^{-\CK(X^{I})} \, \equiv \, i(\bar{X}^I F_I - X^I \bar{F}_I) \, ,
\ee
which for the prepotential~$F$ given in~\eqref{Ffull} 
takes the form:
\be
e^{-\CK(X^{I})} \= \frac{2p^1}{\phi^0}\bigl(m + E_2(\tau) + E_2(-\overline{\t})\bigr) \, . 
\ee
This yields the relation  
\be
M(\t,\overline{\t}) \= \frac1{2^{13}}\frac{\phi^0}{p^1}\;e^{-\CK(\phi^{I})} \, .
\ee

We make a small digression to note that, 
although we work with the specific details of the~$\CN=4$ theory, most of the methods and arguments of this section 
extend to more general~$\CN=2$ theories. Indeed the physical origin of~$\tau$ as the off-shell fluctuations of the 
axion-dilaton (discussed in~\S\ref{locsugra}) motivates the following form for the measure:
\be \label{Mform}
M(\t,\overline{\t}) = \sum_{r,\bar{r}=0}^\infty\,\widetilde{M}(r,\bar{r})\,q^r\,\bar{q}^{\bar{r}} \, ,
\ee
which is indeed satisfied for the~$K3\times T^{2}$ example, and is 
consistent with the fact that~$M$ approaches a constant as~$\t_{2} \to \infty$, consistent with the large-charge limit.

The function~$\CF^{(1)}$ has a Fourier expansion in powers of~$q=e^{2 \pi i \t}$: 
\be \label{defdp}
e^{-\CF^{(1)}(\t)} = \sum_{p=-1}^\infty\,d(p)\,q^p \, ,
\ee
with~$d(p)$ for positive~$p$ being the number of instantons with charge~$p$. 
Combining the measure factor~\eqref{E2etarel},~\eqref{ME2}, we have (with~$N_{0}=2^{-12}$):
\bea \label{expM}
M(\t,\bar{\t})\,e^{-\mathcal{F}^{(1)}(\t) - \mathcal{F}^{(1)}(-\overline{\t})} &=& 
N_{0} \sum_{p,\bar{p}\,=\,-1}^{\infty}\,(m - p -\bar{p})\,d(p)\,d(\bar{p})\,q^p\,\bar{q}^{\bar{p}} \, , \\
&=& N_{0} \sum_{p,\bar{p}\,=\,-1}^{\infty}\,(m - p -\bar{p})\,d(p)\,d(\bar{p})\,e^{2\pi i(p-\bar{p})\t_1}\,e^{-2\pi(p+
\bar{p})\t_2} \, . \nonumber
\eea

We now plug in the expansion~\eqref{expM} in the quantum entropy integral~\eqref{WhatInst}. 
For each term in this series, we can complete the square in~$\t_{1}$ to get a quadratic Gaussian integrand. 
If we perform the~$\t_{1}$ integral naively over the real line, each term in the above series would lead to 
an integral over~$\t_{2}$ of the form~\eqref{intrep}. 
It would seem that we get an infinite series of $I$-Bessel functions for~$\wh{W}(n,\ell,m)$. 
We remind the reader that it is not surprising to find an infinite series of Bessel functions---indeed the discussion 
of~\S\ref{sec:deg} shows that the microscopic degeneracy has the same structure with the Bessel functions 
having successively sub-leading arguments. We find, however, that the arguments of the Bessel functions here 
decrease (as we expect) up to a point, but then increase indefinitely, thus showing that this 
sum is not convergent! 

A solution to this puzzle was presented recently in~\cite{Gomes:2015xcf} by making a choice of 
contour~$\gamma$ in~\eqref{WhatInst} and analyzing the contributions to the degeneracies from 
each term in the Fourier expansion. 
With this choice of contour, almost all of the infinite number of Bessel functions turn out to 
be highly suppressed, and one is left with a finite number of~$I$-Bessel functions, 
consistent with the structure of the leading~$c=1$ term of the Rademacher expansion~\eqref{radi}. 
We now review this analysis, and use the contour prescription of~\cite{Gomes:2015xcf} to 
make a detailed comparison between the expansion of the integral~\eqref{eq:davidsen} and 
the~$c=1$ term of the Rademacher expansion~\eqref{radi}. 
In making this comparison, we provide some clarifications about the details of 
the analysis of~\cite{Gomes:2015xcf}, which we present in Appendix~\ref{app:Iu}. 
We find, at the end of our analysis, that the two expansions actually agree in great detail, 
in the appropriate regime of validity, including the integer coefficients of the Bessel functions!
At first sight this observation may seem to be a pleasant surprise about this particular~$\CN=4$ string theory, but  
as we sketch in the introduction, it can be understood as a reflection of the deeper and broader ideas 
of~\cite{Ooguri:2004zv, Denef:2007vg}, namely that worldsheet instanton degeneracies 
encode the microscopic black hole degeneracies in a very precise manner.

The analysis begins with the expansion~\eqref{expM} in the expression~\eqref{WhatInst}.
Splitting the contour~$\gamma$ into 
two contours~$\gamma_1$,~$\gamma_2$ for the~$\t_1$ and~$\t_2$ integrals, respectively, and 
completing the square in each term, we obtain:
\bea
\label{eq:perfectsquare}
\wh{W}(n,\ell,m) &=& N_0\sum_{p,\bar{p}\,\geq\,-1}(m - p - \bar{p})d(p)d(\bar{p})\,e^{i\pi(p-\bar{p})\frac{\ell}{m}} \, \times \cr
&&\times \; \int_{\gamma_2}\frac{d\tau_2}{\tau_2^{(\nv+3)/2}} \exp\Bigl[-\pi\t_2 \frac{\D(p,\bar{p})}{m} + \frac{\pi}{\t_2}\bigl(n - \frac{\ell^2}{4m}\bigr)\Bigr] \, \times \\
&&\times \; \int_{\gamma_1} d\tau_1 \exp\Biggl[\frac{\pi m}{\tau_2}\Bigl(\t_1 + i(p-\bar{p})\frac{\t_2}{m}-\frac{\ell}{2m}\Bigr)^2\Biggr] \, , \nonumber
\eea
where we have defined
\be \label{defK}
\D(p,\bar{p}) := 4m\bar{p} - (m-(p-\bar{p}))^2 \, .
\ee
(We will see in the following that the function~$\D$ becomes precisely the polar discriminants entering the 
Rademacher expansion~\eqref{radi}.)
We now define the contours~$\gamma_1$,~$\gamma_2$ pertaining to the~$\t_1$ and~$\t_2$ integrals 
as~\cite{Gomes:2015xcf}
\bea \label{contours}
\tau_1 &=& i\tau_2\,u\;\, : \;\; -1+\delta \leq u \leq 1-\delta \, , \cr
\tau_2\,&:&\epsilon - i\infty < \tau_2 < \epsilon + i\infty \, ,
\eea
with~$\delta$ small and positive and~$\epsilon$ positive. This choice ensures that~$|q|<1$ and~$|\bar{q}|<1$ so that 
the Fourier expansion~\eqref{expM} is convergent. Continuing as in~\cite{Gomes:2015xcf}, 
we now define the integral
\be \label{Iu}
I_u(p,\bar{p}) = \int_{\gamma_1} d\tau_1 \exp\Biggl[\frac{\pi m}{\tau_2}\Bigl(\t_1 + i(p-\bar{p})\frac{\t_2}{m}-\frac{\ell}{2m}\Bigr)^2\Biggr] \, .
\ee
Defining~$\alpha \equiv (p-\bar{p})/m$, there are two types of contributions 
to~$\wh{W}(n,\ell,m)$ depending on whether~$|\alpha| \leq 1-\delta$ or~$|\alpha| > 1-\delta$. The leading contributions to the sum~\eqref{eq:perfectsquare} 
are for~$|\alpha| \leq 1-\delta$, and the terms for which~$|\alpha| > 1-\delta$ are exponentially suppressed. 
We then need to take a~$\delta \rightarrow 0$ limit in the contour~$\gamma_1$. This limit is rather subtle, but it can be shown 
that once we take it, the leading contributions to the sum~\eqref{eq:perfectsquare} are the ones for which~$|\alpha| \leq 1$ 
(modulo what we call ``edge-effects'', as we discuss in Appendix~\ref{app:Iu}).

Focusing on these contributions to the quantum entropy, we may evaluate the~$\tau_1$ integral in~\eqref{eq:perfectsquare} and we 
are left with the~$\tau_2$ integral. The latter will yield exponentially growing~$I$-Bessel functions~\eqref{intrep} as long as~$\Delta(p,\bar{p}) < 0$.
Therefore, we now have two conditions,~$|\alpha| \leq 1$ and~$\Delta < 0$, which can be used to bound the sums 
over~$(p,\bar{p})$. Putting these facts together leads to the following expression for~$\wh{W}$:
\bea
\wh{W}(n,\ell,m) &\simeq& N_0\sum_{p,\bar{p}\,\geq\,-1} \; \sum_{\substack{-m \, \leq \, p-\bar{p} \, \leq \, m \\ \Delta(p,\bar{p}) \, < \, 0}}(m - p - \bar{p})\,d(p)\,d(\bar{p})\,e^{i\pi(p-\bar{p})\frac{\ell}{m}} \; \times \cr
&&\times \; \frac{i}{\sqrt{m}}\int_{\gamma_2} \frac{d\tau_2}{\tau_2^{(\nv+2)/2}} \exp\Bigl[-\pi\t_2 \frac{\Delta}{m} + \frac{\pi}{\t_2}\bigl(n-\frac{\ell^2}{4m}\bigr)\Bigr] \, .
\eea
Here the~$\;\simeq\;$ sign is taken to mean that we have thrown away exponentially suppressed contributions to the complete answer for~$\wh{W}$. We can now evaluate the remaining integral on the contour~$\gamma_2$, which yields a Bessel function:
\bea\label{Whatfinal}
\wh{W}(n,\ell,m) &\simeq& N_0\sum_{p,\bar{p}\,\geq\,-1} \; \sum_{\substack{-m \, \leq \, p - \bar{p} \, \leq \, m \\ \Delta(p,\bar{p}) \, < \, 0}}(m - p - \bar{p})\,d(p)\,d(\bar{p})\,e^{i\pi(p-\bar{p})\frac{\ell}{m}} \, \times \cr
&&\times \;\frac{2\pi}{\sqrt{m}}\Biggl(\frac{-\Delta(p,\bar{p})/m}{n-\frac{\ell^2}{4m}}\Biggr)^{\nv/4}\,I_{\nv/2}\Biggl(2\pi\sqrt{-\frac{\Delta(p,\bar{p})}{m}\Bigl(n-\frac{\ell^2}{4m}\Bigr)}\Biggr) \, .
\eea
The symmetry~$\Delta(p,\bar{p}) = \Delta(\bar{p},p)$ implies that one can write the above expression as 
a sum over~$p-\overline{p}$ from~$0$ to~$m$, with the replacement of the phase~$e^{i\pi(p-\bar{p})\frac{\ell}{m}}$
by~$\cos \bigl(\pi(p-\bar{p})\frac{\ell}{m} \bigr)$. 

To proceed further, we make the following change of variables:
\be
\ell' \equiv m-(p-\bar{p}) \, , \qquad n' \equiv \bar{p} \, .
\ee
In these variables, we have~$\D(n',\ell') = 4mn'-\ell'^{2}$ as anticipated, 
and~\eqref{Whatfinal} takes the form
\bea \label{WhatRadvars}
\wh{W}(n,\ell,m) &\simeq& 2N_0\sum_{\substack{0\,\leq\,\ell'\,\leq\,m \\ n'\,\geq\,-1}} \; \sum_{4n' - \frac{\ell'^2}{m} \, < \, 0}(\ell'-2n')\,d(m+n'-\ell')\,d(n')\,\cos{\bigl(\pi(m-\ell')\frac{\ell}{m}\bigr)}\; \times \cr 
&&\times \; \frac{2\pi}{\sqrt{m}} \Biggl(\frac{\bigl|4n' - \frac{\ell'^2}{m}\bigr|}{n-\frac{\ell^2}{4m}}\Biggr)^{\nv/4}\,I_{\nv/2}\Biggl(2\pi\sqrt{\Bigl|4n' - \frac{\ell'^2}{m}\Bigr|\Bigl(n - \frac{\ell^2}{4m}\Bigr)}\Biggr) \, .
\eea
In this form,~$\wh{W}$ can readily be compared to the leading Rademacher expansion for a Jacobi form of index~$m$ and weight~$(3-\nv)/2$. Indeed for such a Jacobi form, the~$c=1$ term of the Rademacher 
expansion~\eqref{radi},~\eqref{Kloosc1} reads
\bea \label{RadJac}
c(n,\ell) &\simeq& \frac{1}{2^{(\nv-1)/2}}\sum_{0 \, \leq \, \ell' \, \leq \, m}\;\sum_{4n'- \tfrac{\ell'^2}{m} \, < \, 0}c(n',\ell')\,\cos{\bigl(\pi(m-\ell')\frac{\ell}{m}\bigr)}\; \times \cr
&&\times \; \frac{2\pi}{\sqrt{m}} \Biggl(\frac{\Bigl|4n' - \frac{\ell'^2}{m}\Bigr|}{n-\frac{\ell^2}{4m}}\Biggr)^{\nv/4} I_{\nv/2}\Biggl(2\pi\sqrt{\Bigl|4n' - \frac{\ell'^2}{m}\Bigr|\Bigl(n - \frac{\ell^2}{4m}\Bigr)}\Biggr) \, .
\eea

We see that~\eqref{WhatRadvars} 
has exactly the same form as~\eqref{RadJac} if we make the identification:
\be  \label{cdrel}
c(n,\ell) = (\ell-2n)\,d(m+n-\ell)\,d(n) \, , \qquad 4mn-\ell^{2} <0, \quad n \geq -1, \; 0 \leq \ell \leq m \, .
\ee
We read this formula as an explicit prediction for the left-hand side which are the polar coefficients~$c^\text{F}(n,\ell)$ 
of the mock Jacobi forms~\eqref{psiF} that control the single-centered black hole degeneracies. 
The coefficients~$d(p)$ of the right hand side are the instanton degeneracies captured by the 
function~$\CF^{(1)}$~\eqref{defdp}
\be
\frac{1}{\eta(\t)^{24}} \= \sum_{n\ge -1} d(n) \, q^{n} \= q^{-1} \+ 24  \+ 324q \+ 3200q^2 \+ 25650q^3 \+ 176256q^4 \+ \cdots \, .  
\ee 
The fact that the instanton degeneracies~$d(n)$ vanish for~$n<-1$ is reflected in the fact that 
the single centered degneracies~$c^\text{F}(n,\ell)$ also vanish for~$n<-1$ as we saw briefly in~\S\ref{sec:deg}. 
In the next section we see that the expansion~\eqref{WhatRadvars} agrees very precisely with the Rademacher-like 
expansion for the Fourier coefficients~$c^\text{F}(n,\ell)$---up to an order where the latter starts to deviate from the 
form~\eqref{RadJac} due to its mock modular nature. 

\section{Polar terms in quarter-BPS black holes in~$\CN=4$ theory \label{microforms}} 

In this section we verify the relation~\eqref{cdrel}
for the first few values of~$m$. We explain that there are three sources of approximations in our derivation which impose
a regime of validity for the comparison of the macroscopic and the microscopic formulas. 
The first source is that we have only 
kept the first~($c=1$) series in the microscopic Rademacher expansion while we should really keep all the terms 
from~$c=1,2,3,\ldots$ The second source, as we explained in~\S\ref{sec:deg}, is that the effects of the shadow of 
the mock modular forms (although small to leading order) can become relevant at a certain sub-leading order.
The third, as we explain in detail in Appendix~\ref{app:Iu}, is what we call ``edge-effects'' in the evaluation of the 
two-dimensional integral which is the result of the localized supergravity path integral. The first source can be
controlled in a fairly straightforward manner but typically this is the smallest effect. 
The second source is an interesting problem in analytic number theory, and the third is a problem for us to better 
define our contour prescription in supergravity. We leave these two problems for the future. 
We now explain these three effects with examples.

We begin with~$m=1$.  We have:
\be
 \psi_1^\text{F}(\tau,z) =  \frac{1}{\eta(\tau)^{24}}(3 E_4(\tau)A(\tau,z)- 648\mathcal{H}_{1}(\tau,z)) \, , \label{m1}\\
\ee
whose Fourier expansion begins as:
\bea
\psi_1^\text{F}(\tau,z) = &&
 (3\z \+ 48\+ 3\z^{-1})q^{-1} \+ (48\z^2 \+ 600\z - 648 \+ 600\z^{-1} \+ 48\z^{-2}) \+ \nonumber \\
&& (3\z^3 - 648\z^2 \+ 25353\z - 50064 \+ 25353\z^{-1} - 648\z^{-2} \+ 3\z^{-3}) \, q \+ \\ 
&& (600\z^3 - 50064\z^2 \+ 561576\z - 1127472 \+ 561576\z^{-1} - 50064\z^{-2} \+ 600\z^{-3})q^2 \+ \cdots \nonumber
\eea
The polar terms are~$(n,\ell)=(-1,1)$, $(-1,0)$, and~$(0,1)$ or equivalently $(\D,\ell)=(-5,1)$, $(-4,0)$, $(-1,1)$.  
The corresponding coefficients~$c_{1}^\text{F}(n,\ell)$ are\footnote{We note that there is a textual error in 
the appendix of~\cite{Bringmann:2012zr}. In the first paragraph of the appendix, it says that the 
coefficients~$c_{m}^{F}(n,\ell)$ of the mock Jacobi forms are presented for the first four values of~$m$, 
while what is really presented is~$d(n,\ell,m)=(-1)^{\ell} c_{m}^{F}(n,\ell)$ to emphasize the positivity 
of those numbers. In particular, the polar coefficients~$c_{m}^{F}(n,\ell)$ (i.e.~with~$4mn-\ell^{2}<0$) 
are strictly positive.}~\cite{Bringmann:2012zr}:
\be \label{polarm1}
c_{1}^\text{F}(-1, 1) = 3 \, , \qquad c_{1}^\text{F}(-1, 0) = 48 \, , \qquad c_{1}^\text{F}(0, 1) = 600 \, . 
\ee
The corresponding combinations of the~$(\ell-2n)\,d(m+n-\ell)\,d(n)$ are:
 \be
(n,\ell)=(-1,1): \; 3 \, , \qquad (n,\ell)=(-1,0): \; 48 \, , \qquad (n,\ell)=(0,1): \; 576 \, . 
\ee
We see that the first two coefficients agree, and the third does not. This is exactly what we expect, 
as we explained at the end of~\S\ref{sec:deg}. 
Indeed we have made an approximation in the Rademacher expansion keeping only the leading~$c=1$ term, 
and we have\footnote{We do the comparison at~$\ell = 0$ for simplicity.} 
\be \label{Wm1}
\frac{\wh{W}(n,0,1)}{4\pi N_0} = 3\Bigl(\frac{5}{n}\Bigr)^{23/4}I_{23/2}\bigl(2\pi\sqrt{5n}\bigr) + 48\Bigl(\frac{4}{n}\Bigr)^{23/4}I_{23/2}\bigl(2\pi\sqrt{4n}\bigr)+ 576\Bigl(\frac{1}{n}\Bigr)^{23/4}I_{23/2}\bigl(2\pi\sqrt{n}\bigr) \, ,
\ee
while the~$c=1$ term of the Rademacher expansion of a Jacobi form with the polar coefficients~\eqref{polarm1} is:
\be \label{m1c1}
\frac{c^\text{F}_{1}(n,0)}{4\pi N_0} = 3\Bigl(\frac{5}{n}\Bigr)^{23/4}I_{23/2}\bigl(2\pi\sqrt{5n}\bigr) + 48\Bigl(\frac{4}{n}\Bigr)^{23/4}I_{23/2}\bigl(2\pi\sqrt{4n}\bigr)+ 600\Bigl(\frac{1}{n}\Bigr)^{23/4}I_{23/2}\bigl(2\pi\sqrt{n}\bigr)\, ,
\ee
with~$N_0 = 2^{-12}$. The~$c=2$ series in the expansion~\eqref{radi} starts with~$I_{23/2}\bigl(2\pi\sqrt{5n/4}\bigr)$  
which is larger than the last term in~\eqref{m1c1}, and therefore we do not expect an agreement
at this order for the last coefficients in~\eqref{Wm1} and~\eqref{m1c1}.
This is one of the issues that we need to be careful about in our comparison. 

Secondly, as we stressed at the end of~\S\ref{sec:deg}, we also need to be careful about the interference of the 
mock nature of the functions~$\psi_{m}^\text{F}$. The first time\footnote{We find experimentally that 
for~$m=1,2$ the two expansions agree even including the mock piece, but we believe this is an accident,
which will be explained if we work out the asymptotic expansion of the corresponding mock Jacobi form in detail. This fact is highlighted by the use of a~${}^*$ in the tables below.}
we see this interference is for~$m=3$, where we have\footnote{Again, we do the comparison at~$\ell = 0$ for simplicity.}:
\bea \label{Wm3}
\frac{\sqrt{3}}{4\pi N_0}\wh W(n,0,3) &=& 5\left(\frac{7}{n}\right)^{23/4}I_{23/2}\left(2\pi\sqrt{7n}\right)+ 96 \left(\frac{16}{3n}\right)^{23/4}I_{23/2}\left(2\pi\sqrt{\frac{16}{3}n}\right) \nonumber \\
&+& 972\left(\frac{13}{3n}\right)^{23/4}I_{23/2}\left(2\pi\sqrt{\frac{13}{3}n}\right) + 6400\left(\frac{4}{n}\right)^{23/4}I_{23/2}\left(2\pi\sqrt{4n}\right) \nonumber \\
&+&1728\left(\frac{3}{n}\right)^{23/4}I_{23/2}\left(2\pi\sqrt{3n}\right) + 15552\left(\frac{4}{3n}\right)^{23/4}I_{23/2}\left(2\pi\sqrt{\frac{4}{3}n}\right) \nonumber \\
&+&76800\left(\frac{1}{3n}\right)^{23/4}I_{23/2}\left(2\pi\sqrt{\frac{1}{3}n}\right)\, .
\eea
Correspondingly, the $c=1$ term of the Rademacher expansion~\eqref{radi} for $m=3$ is:
\bea \label{m3c1}
\frac{\sqrt{3}}{4\pi N_0}c_3^\text{F}(n,0) &=& 5\left(\frac{7}{n}\right)^{23/4}I_{23/2}\left(2\pi\sqrt{7n}\right) + 96 \left(\frac{16}{3n}\right)^{23/4}I_{23/2}\left(2\pi\sqrt{\frac{16}{3}n}\right) \nonumber \\
&+& 972\left(\frac{13}{3n}\right)^{23/4}I_{23/2}\left(2\pi\sqrt{\frac{13}{3}n}\right) + 6404\left(\frac{4}{n}\right)^{23/4}I_{23/2}\left(2\pi\sqrt{4n}\right) \nonumber \\
&+&1728\left(\frac{3}{n}\right)^{23/4}I_{23/2}\left(2\pi\sqrt{3n}\right) + 15600\left(\frac{4}{3n}\right)^{23/4}I_{23/2}\left(2\pi\sqrt{\frac{4}{3}n}\right) \nonumber \\
&+&85176\left(\frac{1}{3n}\right)^{23/4}I_{23/2}\left(2\pi\sqrt{\frac{1}{3}n}\right)\, .
\eea

The~$c=2$ term of the Rademacher expansion starts with~$I_{23/2}\bigl(2\pi\sqrt{7n/4}\bigr)$, and we should ignore terms of that order, i.e. the last two Bessels in~\eqref{m3c1}. However, we still see a disagreement
for the Bessel~$I_{23/2}\left(2\pi\sqrt{4n}\right)$. This is precisely the interference
from the mixed mock Jacobi form~$\v^\text{opt}_{2,3}(\t,z)/\eta(\t)^{24}$. Therefore we 
should only expect agreement up to the Bessel functions~$I_{23/2}(2\pi\sqrt{4n})$. In the 
expressions~\eqref{Wm3},~\eqref{m3c1}, this means that we should not expect a matching of the 
coefficients for the fourth terms,~$6400$ vs.~$6404$.

Thirdly, in deriving our Rademacher-like expression from the supergravity path integral, we made a choice of contour in~\eqref{contours}. 
As explained in Appendix~\ref{app:Iu}, there are ``edge-effects'' in this contour that we have not taken into account properly here. 
These may go towards explaining the boxed discrepancies in the tables below for~$m=5$ and~$m=7$. We believe a more detailed analysis of 
the integral~$I_u(p,\bar{p})$ in~\eqref{Iu} would resolve these discrepancies.

We checked up to~$m=7$ that this kind of an agreement holds exactly after taking into account these three effects. 
We present the data in the form of tables below. 

\paragraph{Legend for tables}:
The pair~$(n,\ell)$ satisifies the conditions in~\eqref{cdrel}, i.e.~$n\geq -1$,~$0\leq\ell\leq m$ and~$(4mn-\ell^2) = \Delta < 0$.
The third column is the coefficient~$c^\text{F}(n,\ell)$ of the mock Jacobi forms~$\psi^\text{F}_{m}$ \eqref{psiF}. Recall that the
black hole exists for positive values of~$\D$ and the degneracy~$c_{m}(n,\ell)$ is controlled by the polar coefficients 
through an expansion of the type~\eqref{RadJac}. 
(Essentially a polar term labelled by~$\D$ enters the analytic formula for the degeneracy~$c_{m}(n,\ell)$ for~$4mn-\ell^{2}>0$ 
at an order~$\exp(2\pi |\D| (4n-\ell^{2}))$.) 
The coefficients below the horizontal line have deviations from their true values because 
we have only included the~$c=1$ series of the Rademacher expansion, while 
at these orders we should necessarily start including the~$c\ge 2$ series. 
We indicate in bold face when the Rademacher expansion cannot be trusted because we have treated a 
mock Jacobi form as a true Jacobi form. (For~$m=1,2$ the coefficient still agree---which we indicate by a~$\bf{}^{*}$.)
As we see clearly in the tables, the deviations for the bold-faced coefficients are small and should be resolved 
by including the effects of the shadow. The boxed values indicate possible edge-effects in the contour prescription. 

\vspace{1cm}

{$\bf m=1$:}

\begin{center}
\begin{tabular}{|c|c|c|c|}
\hline
$\Delta$ & $(n,\ell)$ & $c_1(n,\ell)$ & $(\ell-2n)\,d(1+n-\ell)\,d(n)$ \\
\hline
$-5$ & $(-1,1)$ & 3 & 3 \\
$-4$ & $(-1,0)$ & \bf{48}* & 48 \\
\hline
$-1$ & $(0,1)$ &  {600} & 576 \\
\hline
\end{tabular}
\end{center}

\vspace{1cm}

{$\bf m=2$:}

\begin{center}
\begin{tabular}{|c|c|c|c|}
\hline
$\Delta$ & $(n,\ell)$ & $c_2(n,\ell)$ & $(\ell-2n)\,d(2+n-\ell)\,d(n)$ \\
\hline
$-12$ & $(-1,2)$ & 4 & 4 \\
$-9$ & $(-1,1)$ & 72 & 72 \\
$-8$ & $(-1,0)$ & \bf{648}* & 648 \\
$-4$ & $(0,2)$ & 1152 & 1152 \\
\hline
$-1$ & $(0,1)$ &  {8376} &  7776 \\
\hline
\end{tabular}
\end{center}

\vspace{1cm}

{$\bf m=3$:}

\begin{center}
\begin{tabular}{|c|c|c|c|}
\hline
$\Delta$ & $(n,\ell)$ & $c_3(n,\ell)$ & $(\ell-2n)\,d(3+n-\ell)\,d(n)$ \\
\hline
$-21$ & $(-1,3)$ & 5 & 5 \\
$-16$ & $(-1,2)$ & 96 & 96 \\
$-13$ & $(-1,1)$ & 972 &  972 \\
$-12$ & $(-1,0)$ & \bf{6404} & 6400 \\
$-9$ & $(0,3)$ & 1728 & 1728 \\
\hline
$-4$ & $(0,2)$ &  {15600} & 15552 \\
$-1$ & $(0,1)$ &  {85176} &  76800 \\
\hline
\end{tabular}
\end{center}

\vspace*{\fill}

\newpage

\vspace*{\fill}

{$\bf m=4$:}

\begin{center}
\begin{tabular}{|c|c|c|c|}
\hline
$\Delta$ & $(n,\ell)$ & $c_4(n,\ell)$ & $(\ell-2n)\,d(4+n-\ell)\,d(n)$ \\
\hline
$-32$ & $(-1,4)$ & 6 & 6 \\
$-25$ & $(-1,3)$ & 120 & 120 \\
$-20$ & $(-1,2)$ & 1296 &  1296 \\
$-17$ & $(-1,1)$ & 9600 & 9600 \\
$-16$ & $(0,4)$ & 2304 & 2304 \\
$-16$ & $(-1,0)$ & \bf{51396} & 51300 \\
$-9$ & $(0,3)$ & 23328 & 23328 \\
\hline
$-4$ & $(0,2)$ &  {154752} & 153600 \\
$-1$ & $(0,1)$ &  {700776} &  615600 \\
\hline
\end{tabular}
\end{center}

\vspace{1cm}

{$\bf m=5$:}

\begin{center}
\begin{tabular}{|c|c|c|c|}
\hline
$\Delta$ & $(n,\ell)$ & $c_5(n,\ell)$ & $(\ell-2n)\,d(5+n-\ell)\,d(n)$ \\
\hline
$-45$ & $(-1,5)$ & 7 & 7 \\
$-36$ & $(-1,4)$ & 144 & 144 \\
$-29$ & $(-1,3)$ & 1620 &  1620 \\
$-25$ & $(0,5)$ & 2880 & 2880 \\
$-24$ & $(-1,2)$ & 12800 & 12800 \\
$-21$ & $(-1,1)$ & 76955 & \boxed{76950}  \\
$-20$ & $(-1,0)$ & \bf{353808} & 352512 \\
$-16$ & $(0,4)$ & 31104 & 31104 \\
\hline
$-9$ & $(0,3)$ &  {230472} & 230400 \\
$-5$ & $(1,5)$ &  {315255} & 314928 \\
$-4$ & $(0,2)$ &  {1246800} & 1231200 \\
$-1$ & $(0,1)$ &  {4930920} & 4230144 \\
\hline
\end{tabular}
\end{center}

\vspace*{\fill}

\newpage

\vspace*{\fill}

{$\bf m=6$:}

\begin{center}
\begin{tabular}{|c|c|c|c|}
\hline
$\Delta$ & $(n,\ell)$ & $c_6(n,\ell)$ & $(\ell-2n)\,d(6+n-\ell)\,d(n)$ \\
\hline
$-60$ & $(-1,6)$ & 8 & 8 \\
$-49$ & $(-1,5)$ & 168 & 168 \\
$-40$ & $(-1,4)$ & 1944 &  1944 \\
$-36$ & $(0,6)$ & 3456 & 3456 \\
$-33$ & $(-1,3)$ & 16000 & 16000 \\
$-28$ & $(-1,2)$ & 102600 & 102600 \\
$-25$ & $(0,5)$ & 38880 & 38880 \\
$-25$ & $(-1,1)$ & \bf{528888} & 528768 \\
$-24$ & $(-1,0)$ & \bf{2160240} & 2147440 \\
$-16$ & $(0,4)$ & 307200 & 307200 \\
\hline
$-12$ & $(1,6)$ &  {419904} &  419904 \\
$-9$ & $(0,3)$ &  {1848528} & 1846800 \\
$-4$ & $(0,2)$ &  {8615040} & 8460288 \\
$-1$ & $(0,1)$ &  {30700200} & 25769280 \\
$-1$ & $(1,5)$ &  {3118776} & 3110400 \\
\hline
\end{tabular}
\end{center}

\vspace{1cm}

{$\bf m=7$:}

\begin{center}
\begin{tabular}{|c|c|c|c|}
\hline
$\Delta$ & $(n,\ell)$ & $c_7(n,\ell)$ & $(\ell-2n)\,d(7+n-\ell)\,d(n)$ \\
\hline
$-77$ & $(-1,7)$ & 9 & 9 \\
$-64$ & $(-1,6)$ & 192 & 192 \\
$-53$ & $(-1,5)$ & 2268 &  2268 \\
$-49$ & $(0,7)$ & 4032 & 4032 \\
$-44$ & $(-1,4)$ & 19200 & 19200 \\
$-37$ & $(-1,3)$ & 128250 &  128250 \\
$-36$ & $(0,6)$ & 46656 & 46656 \\
$-32$ & $(-1,2)$ & 705030 & \boxed{705024} \\
$-29$ & $(-1,1)$ & 3222780 & \boxed{3221160} \\
$-28$ & $(-1,0)$ & \bf{11963592} & 11860992 \\
$-25$ & $(0,5)$ & 384000 & 384000 \\
$-21$ & $(1,7)$ & 524880 & 524880 \\
\hline
$-16$ & $(0,4)$ &  {2462496} & 2462400 \\
$-9$ & $(0,3)$ &  {12713760} & 12690432 \\
$-8$ & $(1,6)$ &  {4147848} & 4147200 \\
$-4$ & $(0,2)$ &  {52785360} & 51538560 \\
$-1$ & $(0,1)$ &  {173032104} & 142331904 \\
\hline
\end{tabular}
\end{center}

\vspace*{\fill}

\vspace{1cm}

\section*{Acknowledgements}
We thank Satoshi Nawata for initial collaboration on this project. 
We thank Atish Dabholkar, Jo\~ao Gomes, Rajesh Gupta, and Bernard de Wit for useful conversations. 
This work is supported by the EPSRC First Grant UK EP/M018903/1 and the  
ERC Advanced Grant no.~246974, {\it ``Supersymmetry: a window to non-perturbative physics''}.

\vspace{0.4cm}

\appendix

\section{Single-center degeneracies and mock Jacobi forms \label{mockapp}}

In this appendix we briefly review some facts from~\cite{Dabholkar:2012nd} that are relevant for the discussion
of \S\ref{sec:deg} and \S\ref{microforms}.
First we recall the construction of the single-centered black hole partition functions from the Igusa cusp form.
The first step in~\cite{Dabholkar:2012nd} to analyze the single-center Fourier coefficients is to expand the 
microscopic partition function in $e^{2 \pi i \s}$:
\be\label{reciproigusa}
  \frac 1{\Phi_{10}(\t, z, \sigma)} \= \sum_{m\geq -1} \psi_m (\t,z) \, e^{2 \pi i m \s}  \, .
\ee
One then defines the polar part of $\psi_{m}$ (with $q=e^{2\pi i \t}$, $\zeta = e^{2 \pi i z}$)
\be\label{Tm}
\psi^{\text{P}}_m (\t, z) := \; \frac{p_{24}(m+1) }{\eta^{24}(\t)} \, \sum_{s\in\IZ} \, \frac{q^{ms^2 +s}\zeta^{2ms+1}}{(1 -\zeta q^s )^2} \, , 
\ee
where $p_{24}(n)$ counts the number of  partitions of an integer $n$ with $24$ colors.
The function~$\psi^{\text{P}}_{m}$ is the average over the lattice $ \mathbb{Z} \t + \IZ$ of the leading behavior of the function
near the pole $z=0$
  \be \label{simplewall}  
  \frac{p_{24}(m+1)}{\eta(\t)^{24}} \frac{\zeta}{(1-\zeta)^2}\,. \ee
The function $\psi_{m}^{\text{P}}$ is an example of an Appell-Lerch sum, and it encodes the physics of all the wall-crossings due to the decay of two-centered black holes.

The two functions $\psi_{m}$ and $\psi_{m}^{\text{P}}$ have, by
construction, the same poles and residues, so the difference 
\be  
\psi_{m}^\text{F} := \psi_{m} - \psi_{m}^\text{P} \, ,
\ee
called the \emph{finite} or Fourier part of $\psi_{m}$, is holomorphic in $z$, and has an unambiguous Fourier expansion:
\be \label{psiF2}
\psi_{m}^\text{F}(\t,z) \= \sum_{n,\ell} \, c^{\text{F}}_{m}(n,\ell) \, q^{n} \, \zeta^{\ell} \, .
\ee
The indexed degeneracies of the single-centered black hole of magnetic charge invariant~$Q_{m}^{2}/2=m$,
as defined by the attractor mechanism, are related to the Fourier coefficients of the function~$\psi_{m}^{\text{F}}$
as $d(n,\ell,m) = (-1)^{\ell+1} c^{\text{F}}_{m}(n,\ell)$,
the overall sign coming from an analysis of the fermion zero modes described in~\cite{Dabholkar:2010rm}.
The statement of the main theorem of~(\cite{Dabholkar:2012nd}, Chapter 8) is that 
the single-center black hole partition function~$\psi^\text{F}_{m}(\t,z)$ is a \emph{mock Jacobi form}.

What this means is that~$\psi^\text{F}$ has the same elliptic transformation property~\eqref{elliptic} 
as a regular Jacobi form governed by the index~$m$. Its modular transformation property~\eqref{modtransform}, however, 
is modified. The lack of modularity is governed by the explicit function called the~\emph{shadow}:
\be
\psi^\text{S} (\t,z) \=  \frac{1}{\eta(\t)^{24}} \sum_{\ell\in \IZ/2m\IZ} \vth^{*}_{m,\ell}(\t,0) \, \vth_{m,\ell}(\t,z) \, ,
\ee
where the operation~$*$ is such that, a modular form~$g$ of weight~$w$ obeys 
  \be\label{ddtbarh}   (4\pi\t_2)^w\,\,\frac{\partial g^{*}(\t)}{\partial \overline{\tau}} \= -2\pi i\;\overline{g(\tau)}\;.  \ee
The function
\be
\widehat{\psi^\text{F}}(\t,z) = \psi^\text{F} (\t,z) + \psi^{\text{S}} (\t,z) \, , 
\ee 
called the \emph{completion} of~$\psi^\text{F}$, transforms as a Jacobi form of weight~$-10$ and index~$m$, but it is not holomorphic. 
It obeys the holomorphic anomaly equation:
\be\label{ddtbarhphi}   
 (4\pi\t_2)^{1/2} \,\,\frac{\partial \widehat{\psi^\text{F}} (\t,z)}{\partial \overline{\tau}} \= 
  -2\pi i\; \frac{1}{\eta(\t)^{24}} \sum_{\ell\in \IZ/2m\IZ} \overline{\vth_{m,\ell}(\t,0)} \, \vth_{m,\ell}(\t,z) \, . 
\ee

\vspace{0.4cm}

Now we briefly present some relevant facts about the growth of the coefficients of the mock Jacobi forms~$\psi^\text{F}_{m}$. 
By multiplying~$\psi^\text{F}_{m}$ by the function~$\eta(\t)^{24}$, we get a 
function~$\v_{2,m}^\text{mock} = \eta^{24} \psi^\text{F}_{m}$ 
which is a mock Jacobi form of weight~2 and index~$m$.  
It was shown in~(\cite{Dabholkar:2012nd}, Chapters~9,~10) that~$\v_{2,m}$ can be 
written\footnote{Recall that the definition of a mock Jacobi form only holds modulo the addition of a true Jacobi form.} 
as a linear combination of a (true) weak Jacobi form and a mock Jacobi form 
\be
\v^\text{mock}_{2,m}(\t,z) \= \v^\text{true}_{2,m}(\t,z) + \v^\text{opt}_{2,m}(\t,z) \, , 
\ee
such that the mock Jacobi form~$\v^\text{opt}_{2,m}$ has \emph{optimal growth}. This means that the 
Fourier-Jacobi coefficients of~$\v^\text{opt}_{2,m}(\t,z)$ grow at most as
\be \label{optgrowth}
c^\text{opt}(n,\ell) \sim \exp\bigl(\frac{\pi}{m}\sqrt{4mn-\ell^{2}} \bigr)  \, .
\ee
If we look at the Rademacher expansion~\eqref{radi}, the growth~\eqref{optgrowth} is the smallest possible one, governed by 
the value of~$|\wt\D|=1$. In fact, for~$m$ a prime power, the coefficients of the optimal mock Jacobi form has only polynomial 
growth. 

\section{The~$I_u$ integral \label{app:Iu}}

In this appendix, we perform a detailed analysis of the~$I_u$ integral defined in~\eqref{Iu}. 
On the contour~$\gamma_1$ defined in~\eqref{contours}, we have:
\bea
I_u(p,\bar{p}) &=& i\tau_2\int_{-1+\delta}^{1-\delta}\,du\,\exp\Bigl[-\pi m\tau_2(u + \alpha - \frac{\ell}{2im\tau_2})^2\Bigr] \, , \\
&=& \frac{1}{2}\sqrt{\frac{\tau_2}{m}}\Bigl[\text{Erfi}\bigl(\frac{\sqrt{\pi}(\ell-2i\tau_2 m(\alpha-1+\delta)}{2\sqrt{\tau_2 m}}\bigr) - \text{Erfi}\bigl(\frac{\sqrt{\pi}(\ell-2i\tau_2 m(\alpha+1-\delta)}{2\sqrt{\tau_2 m}}\bigr)\Bigr] \, , \nonumber
\eea
where~$\text{Erfi}(x)$ is the imaginary error function~$\text{Erfi}(x) = \text{Erf}(ix)/i$. Above we have defined~$\alpha \equiv (p-\bar{p})/m$, and~$\delta$ is a small, positive constant parametrizing the contour~$\gamma_1$.

We now take~$\text{Re}(\tau_2) = \epsilon$ to be very large and use the Taylor series of the imaginary error function in this regime. Depending on the value of~$|\alpha|$, we obtain three results: first for~$|\alpha| < 1-\delta$, where the Taylor expansion gives:
\bea \label{Iualphaless}
I^{|\alpha|< 1-\delta}_u = i\sqrt{\frac{\tau_2}{m}} &+& \exp\Bigl[\pi\frac{(\ell-2i\tau_2 m(\alpha-1+\delta))^2}{4\tau_2 m}\Bigr]\Bigl(\frac{i}{2\pi m(\alpha-1+\delta)} + \mathcal{O}\bigl(\frac{1}{\epsilon}\bigr)\Bigr) \cr
&-& \exp\Bigl[\pi\frac{(\ell-2i\tau_2 m(\alpha+1-\delta))^2}{4\tau_2 m}\Bigr]\Bigl(\frac{i}{2\pi m(\alpha+1-\delta)} + \mathcal{O}\bigl(\frac{1}{\epsilon}\bigr)\Bigr) \, .
\eea
Secondly, for~$|\alpha| = 1-\delta$:
\bea \label{Iualphaequal}
I^{\alpha = \pm(1-\delta)}_u &=& \frac{i}{2}\sqrt{\frac{\tau_2}{m}} \pm \frac{\ell}{2m} \cr
&&-\exp\Bigl[\pi\frac{(\ell\pm4i\tau_2 m(\delta-1))^2}{4\tau_2 m}\Bigr]\Bigl(\frac{i}{4\pi m(1-\delta)} + \mathcal{O}\bigl(\frac{1}{\epsilon}\bigr)\Bigr) \, .
\eea
Lastly, for~$|\alpha| > 1-\delta$:
\bea \label{Iualphamore}
I^{|\alpha| > 1-\delta}_u &=& \exp\Bigl[\pi\frac{(\ell-2i\tau_2 m(\alpha-1+\delta))^2}{4\tau_2 m}\Bigr]\Bigl(\frac{i}{2\pi m(\alpha-1+\delta)} + \mathcal{O}\bigl(\frac{1}{\epsilon}\bigr)\Bigr) \cr
&&-\exp\Bigl[\pi\frac{(\ell-2i\tau_2 m(\alpha+1-\delta))^2}{4\tau_2 m}\Bigr]\Bigl(\frac{i}{2\pi m(\alpha+1-\delta)} + \mathcal{O}\bigl(\frac{1}{\epsilon}\bigr)\Bigr) \, .
\eea
Focusing on the case where~$|\alpha| < 1-\delta$, we can use the above expressions for~$I_u$ in~\eqref{eq:perfectsquare}. The~$\tau_2$ integral is now on a contour where~$\epsilon \gg 1$, but since the only pole in the~$\tau_2$ complex plane sits at the origin, we can safely deform it back to~$\epsilon$ small and still positive. On this contour, we find using~\eqref{intrep}:
\bea
\wh{W}^{|\alpha| < 1-\delta} = N_0 \sum_{p,\bar{p}\geq-1} \Biggl[&&\sum_{\substack{-m(1-\delta) < p - \bar{p} < m(1-\delta) \\ \Delta(p,\bar{p}) < 0}}(m - p - \bar{p})\,d(p)\,d(\bar{p})\,e^{i\pi(p-\bar{p})\frac{\ell}{m}} \; \times \cr
&&\times \;\frac{2\pi}{\sqrt{m}}\Biggl(\frac{|\Delta(p,\bar{p})/m|}{n-\frac{\ell^2}{4m}}\Biggr)^{\nv/4}\,I_{\nv/2}\Biggl(2\pi\sqrt{\left|\frac{\Delta(p,\bar{p})}{m}\right|\Bigl(n-\frac{\ell^2}{4m}\Bigr)}\Biggr) \cr
+&&\sum_{\substack{-m(1-\delta) < p - \bar{p} < m(1-\delta) \\ 4\bar{p}-2\delta(\bar{p}-p+m)+m\delta^2 < 0}}(m - p - \bar{p})\,d(p)\,d(\bar{p})\,e^{i\pi\ell(1-\delta)} \; \times \cr
&&\times \;\frac{1}{p-\bar{p}-m(1-\delta)}\Biggl(\frac{|4\bar{p}-2\delta(\bar{p}-p+m)+m\delta^2|}{n}\Biggr)^{(\nv+1)/4} \; \times \cr
&&\times \;I_{(\nv+1)/2}\Biggl(2\pi\sqrt{|4\bar{p}-2\delta(\bar{p}-p+m)+m\delta^2|n}\Biggr) \cr
+&&\sum_{\substack{-m(1-\delta) < p - \bar{p} < m(1-\delta) \\ 4p-2\delta(p-\bar{p}+m)+m\delta^2 < 0}}(m - p - \bar{p})\,d(p)\,d(\bar{p})\,e^{-i\pi\ell(1-\delta)} \; \times \cr
&&\times \;\frac{1}{\bar{p}-p-m(1-\delta)}\Biggl(\frac{|4p-2\delta(p-\bar{p}+m)+m\delta^2|}{n}\Biggr)^{(\nv+1)/4} \; \times \cr
&&\times \;I_{(\nv+1)/2}\Biggl(2\pi\sqrt{|4p-2\delta(p-\bar{p}+m)+m\delta^2|n}\Biggr)\Biggr] \, .
\eea
The case~$|\alpha| = 1-\delta$ contributes to~$\wh{W}$ as:
\bea
\wh{W}^{\alpha = \pm(1-\delta)} = N_0 \sum_{p\geq-1} \Biggl[&&\sum_{4mp - m^2\delta^2 < 0}(m\delta - 2p)\,d(p)\,d(p + m - m\delta)\,e^{i\pi(1-\delta)\ell} \; \times \cr
&&\times \;\frac{\pi}{\sqrt{m}}\Biggl(\frac{|4p - m\delta^2|}{n-\frac{\ell^2}{4m}}\Biggr)^{\nv/4}\,I_{\nv/2}\Biggl(2\pi\sqrt{\left|4p - m\delta^2\right|\Bigl(n-\frac{\ell^2}{4m}\Bigr)}\Biggr) \cr
\pm&&\frac{\ell}{2m}\sum_{4mp - m^2\delta^2 < 0}(m\delta - 2p)\,d(p)\,d(p + m - m\delta)\,e^{i\pi(1-\delta)\ell} \; \times \cr
&&\times \;2\pi\Biggl(\frac{|4p - m\delta^2|}{n}\Biggr)^{(\nv+1)/4}I_{(\nv+1)/2}\Biggl(2\pi\sqrt{|4p - m\delta^2|n}\Biggr) \cr
+&&\sum_{4p+4m-8m\delta+3m\delta^2 < 0}(m\delta - 2p)\,d(p)\,d(p + m - m\delta)\,e^{\mp i\pi(1-\delta)\ell} \; \times \cr
&&\times \;\frac{1}{2m(\delta-1)}\Biggl(\frac{|4p+4m-8m\delta+3m\delta^2|}{n}\Biggr)^{(\nv+1)/4} \; \times \cr
&&\times \;I_{(\nv+1)/2}\Biggl(2\pi\sqrt{|4p+4m-8m\delta+3m\delta^2|n}\Biggr)\Biggr] \, .
\eea
At this point, we can put the three contributions together to obtain~$\wh{W}$ for~$|\alpha|\leq 1 -\delta$:
\bea
\wh{W}^{|\alpha| \leq 1-\delta} &=& N_0 \sum_{p,\bar{p}\geq-1} \sum_{\substack{-m(1-\delta) \leq p - \bar{p} \leq m(1-\delta) \\ \Delta(p,\bar{p}) < 0}}(m - p - \bar{p})\,d(p)\,d(\bar{p})\,e^{i\pi(p-\bar{p})\frac{\ell}{m}} \; \times \cr
&&\times \;\frac{2\pi}{\sqrt{m}}\Biggl(\frac{|\Delta(p,\bar{p})/m|}{n-\frac{\ell^2}{4m}}\Biggr)^{\nv/4}\,I_{\nv/2}\Biggl(2\pi\sqrt{\left|\frac{\Delta(p,\bar{p})}{m}\right|\Bigl(n-\frac{\ell^2}{4m}\Bigr)}\Biggr) \cr
+&&N_0 \sum_{p,\bar{p}\geq-1} \sum_{\substack{-m(1-\delta) < p - \bar{p} < m(1-\delta) \\ 4\bar{p}-2\delta(\bar{p}-p+m)+m\delta^2 < 0}}(m - p - \bar{p})\,d(p)\,d(\bar{p})\,e^{i\pi\ell(1-\delta)} \; \times \cr
&&\times \;\frac{1}{p-\bar{p}-m(1-\delta)}\Biggl(\frac{|4\bar{p}-2\delta(\bar{p}-p+m)+m\delta^2|}{n}\Biggr)^{(\nv+1)/4} \; \times \cr
&&\times \;I_{(\nv+1)/2}\Biggl(2\pi\sqrt{|4\bar{p}-2\delta(\bar{p}-p+m)+m\delta^2|n}\Biggr) \cr
+&&N_0 \sum_{p,\bar{p}\geq-1} \sum_{\substack{-m(1-\delta) < p - \bar{p} < m(1-\delta) \\ 4p-2\delta(p-\bar{p}+m)+m\delta^2 < 0}}(m - p - \bar{p})\,d(p)\,d(\bar{p})\,e^{-i\pi\ell(1-\delta)} \; \times \cr
&&\times \;\frac{1}{\bar{p}-p-m(1-\delta)}\Biggl(\frac{|4p-2\delta(p-\bar{p}+m)+m\delta^2|}{n}\Biggr)^{(\nv+1)/4} \; \times \cr
&&\times \;I_{(\nv+1)/2}\Biggl(2\pi\sqrt{|4p-2\delta(p-\bar{p}+m)+m\delta^2|n}\Biggr) \cr
+&&N_0 \sum_{p\geq-1} \sum_{4p+4m-8m\delta+3m\delta^2 < 0}(m\delta - 2p)\,d(p)\,d(p + m - m\delta) \; \times \cr
&&\times \;\frac{\cos\bigl(\pi(\delta-1)\ell\bigr)}{m(\delta-1)}\Biggl(\frac{|4p+4m-8m\delta+3m\delta^2|}{n}\Biggr)^{(\nv+1)/4} \; \times \cr
&&\times \;I_{(\nv+1)/2}\Biggl(2\pi\sqrt{|4p+4m-8m\delta+3m\delta^2|n}\Biggr) \, .
\eea
The remaining contributions to~$\wh{W}$ are the ones for which~$|\alpha| > 1-\delta$. Comparing~\eqref{Iualphaless} and~\eqref{Iualphamore}, it is clear that they will give the same type of contributions as the second and third sums of Bessel functions above, albeit with a different range for~$p,\bar{p}$. Putting everything together, we find the following expression:
\bea \label{app:Whatcomplete}
\wh{W}(n,\ell,m) &=& N_0 \sum_{p,\bar{p}\geq-1} \sum_{\substack{-m(1-\delta) \leq p - \bar{p} \leq m(1-\delta) \\ \Delta(p,\bar{p}) < 0}}(m - p - \bar{p})\,d(p)\,d(\bar{p})\,e^{i\pi(p-\bar{p})\frac{\ell}{m}} \; \times \cr
&&\times \;\frac{2\pi}{\sqrt{m}}\Biggl(\frac{|\Delta(p,\bar{p})/m|}{n-\frac{\ell^2}{4m}}\Biggr)^{\nv/4}\,I_{\nv/2}\Biggl(2\pi\sqrt{\left|\frac{\Delta(p,\bar{p})}{m}\right|\Bigl(n-\frac{\ell^2}{4m}\Bigr)}\Biggr) \cr
+&&N_0 \sum_{p\geq-1} \sum_{4p+4m-8m\delta+3m\delta^2 < 0}(m\delta - 2p)\,d(p)\,d(p + m - m\delta) \; \times \cr
&&\times \;\frac{\cos\bigl(\pi(\delta-1)\ell\bigr)}{m(\delta-1)}\Biggl(\frac{|4p+4m-8m\delta+3m\delta^2|}{n}\Biggr)^{(\nv+1)/4} \; \times \cr
&&\times \;I_{(\nv+1)/2}\Biggl(2\pi\sqrt{|4p+4m-8m\delta+3m\delta^2|n}\Biggr) \cr
+&&N_0 \sum_{p,\bar{p}\geq-1} \sum_{\substack{p-\bar{p}\,\neq\,\pm m(1-\delta) \\ 4\bar{p}-2\delta(\bar{p}-p+m)+m\delta^2 < 0}}(m - p - \bar{p})\,d(p)\,d(\bar{p}) \; \times \cr
&&\times \;\frac{2\cos\bigl(\pi(\delta-1)\ell\bigr)}{p-\bar{p}-m(1-\delta)}\Biggl(\frac{|4\bar{p}-2\delta(\bar{p}-p+m)+m\delta^2|}{n}\Biggr)^{(\nv+1)/4} \; \times \cr
&&\times \;I_{(\nv+1)/2}\Biggl(2\pi\sqrt{|4\bar{p}-2\delta(\bar{p}-p+m)+m\delta^2|n}\Biggr) \, .
\eea
In~\S\ref{ContourPres} (and in~\cite{Gomes:2015xcf}), it was argued that, in the limit where~$\delta \rightarrow 0$, only the first sum of Bessel functions (the ones with weight~$\nv/2$) contribute to~$\wh{W}(n,\ell,m)$ and that the second and third contributions to~\eqref{app:Whatcomplete} are all exponentially suppressed. Indeed, the first term is unambiguous when taking the limit~$\delta \rightarrow 0$, and yields precisely the expression~\eqref{Whatfinal}. However, this limit is less trivial for the Bessel functions of integer weight~$(\nv+1)/2$. In fact, the contribution from these Bessels depends \emph{sensitively} on how~$\delta$ goes to zero, and on the behavior of the product~$m\delta$ in this limit. Since~$\delta$ is a parameter introduced for the contour~$\gamma_1$, this means that a particular choice of~$\gamma_1$ can pick up additional contributions from the integer-weight Bessel functions, which may then become comparable to the Bessels of half-integer weight. This is what we call ``edge-effects'', arising from the limit~$\delta \rightarrow 0$ in the contour~$\gamma_1$~\eqref{contours}. 

\providecommand{\href}[2]{#2}\begingroup\raggedright\endgroup


\begin{thebibliography}{10}
\bibliographystyle{JHEP}

\bibitem{Sen:2008vm}
A.~Sen, {\it {Quantum Entropy Function from AdS(2)/CFT(1) Correspondence}},
  {\em Int. J. Mod. Phys.} {\bf A24} (2009) 4225--4244,
  [\href{http://arxiv.org/abs/0809.3304}{{\tt arXiv:0809.3304}}].

\bibitem{Pestun:2007rz}
V.~Pestun, {\it {Localization of gauge theory on a four-sphere and
  supersymmetric Wilson loops}},  {\em Commun. Math. Phys.} {\bf 313} (2012)
  71--129, [\href{http://arxiv.org/abs/0712.2824}{{\tt arXiv:0712.2824}}].

\bibitem{Banerjee:2009af}
N.~Banerjee, S.~Banerjee, R.~K. Gupta, I.~Mandal, and A.~Sen, {\it
  {Supersymmetry, Localization and Quantum Entropy Function}},  {\em JHEP} {\bf
  02} (2010) 091, [\href{http://arxiv.org/abs/0905.2686}{{\tt
  arXiv:0905.2686}}].

\bibitem{Dabholkar:2010uh}
A.~Dabholkar, J.~Gomes, and S.~Murthy, {\it {Quantum black holes, localization
  and the topological string}},  {\em JHEP} {\bf 06} (2011) 019,
  [\href{http://arxiv.org/abs/1012.0265}{{\tt arXiv:1012.0265}}].

\bibitem{Dabholkar:2011ec}
A.~Dabholkar, J.~Gomes, and S.~Murthy, {\it {Localization \& Exact
  Holography}},  {\em JHEP} {\bf 04} (2013) 062,
  [\href{http://arxiv.org/abs/1111.1161}{{\tt arXiv:1111.1161}}].

\bibitem{Murthy:2015yfa}
S.~Murthy and V.~Reys, {\it {Functional determinants, index theorems, and exact
  quantum black hole entropy}}, [\href{http://arxiv.org/abs/1504.01400}{{\tt
  arXiv:1504.01400}}].

\bibitem{LopesCardoso:1998wt}
G.~Lopes~Cardoso, B.~de~Wit, and T.~Mohaupt, {\it {Corrections to macroscopic
  supersymmetric black hole entropy}},  {\em Phys. Lett.} {\bf B451} (1999)
  309--316, [\href{http://arxiv.org/abs/hep-th/9812082}{{\tt hep-th/9812082}}].

\bibitem{LopesCardoso:2000qm}
G.~Lopes~Cardoso, B.~de~Wit, J.~Kappeli, and T.~Mohaupt, {\it {Stationary BPS
  solutions in N=2 supergravity with $R^2$ interactions}},  {\em JHEP} {\bf 12}
  (2000) 019, [\href{http://arxiv.org/abs/hep-th/0009234}{{\tt
  hep-th/0009234}}].

\bibitem{Ooguri:2004zv}
H.~Ooguri, A.~Strominger, and C.~Vafa, {\it {Black hole attractors and the
  topological string}},  {\em Phys. Rev.} {\bf D70} (2004) 106007,
  [\href{http://arxiv.org/abs/hep-th/0405146}{{\tt hep-th/0405146}}].

\bibitem{Sen:2005wa}
A.~Sen, {\it {Black hole entropy function and the attractor mechanism in higher
  derivative gravity}},  {\em JHEP} {\bf 09} (2005) 038,
  [\href{http://arxiv.org/abs/hep-th/0506177}{{\tt hep-th/0506177}}].

\bibitem{Castro:2008ys}
A.~Castro and S.~Murthy, {\it {Corrections to the statistical entropy of five
  dimensional black holes}},  {\em JHEP} {\bf 06} (2009) 024,
  [\href{http://arxiv.org/abs/0807.0237}{{\tt arXiv:0807.0237}}].

\bibitem{Dabholkar:2005dt}
A.~Dabholkar, F.~Denef, G.~W. Moore, and B.~Pioline, {\it {Precision counting
  of small black holes}},  {\em JHEP} {\bf 10} (2005) 096,
  [\href{http://arxiv.org/abs/hep-th/0507014}{{\tt hep-th/0507014}}].

\bibitem{Shih:2005he}
D.~Shih and X.~Yin, {\it {Exact black hole degeneracies and the topological
  string}},  {\em JHEP} {\bf 04} (2006) 034,
  [\href{http://arxiv.org/abs/hep-th/0508174}{{\tt hep-th/0508174}}].

\bibitem{Cardoso:2006xz}
G.~L. Cardoso, B.~de~Wit, and S.~Mahapatra, {\it {Black hole entropy functions
  and attractor equations}},  {\em JHEP} {\bf 03} (2007) 085,
  [\href{http://arxiv.org/abs/hep-th/0612225}{{\tt hep-th/0612225}}].

\bibitem{Denef:2007vg}
F.~Denef and G.~W. Moore, {\it {Split states, entropy enigmas, holes and
  halos}},  {\em JHEP} {\bf 11} (2011) 129,
  [\href{http://arxiv.org/abs/hep-th/0702146}{{\tt hep-th/0702146}}].

\bibitem{Banerjee:2008ky}
N.~Banerjee, D.~P. Jatkar, and A.~Sen, {\it {Asymptotic Expansion of the N=4
  Dyon Degeneracy}},  {\em JHEP} {\bf 05} (2009) 121,
  [\href{http://arxiv.org/abs/0810.3472}{{\tt arXiv:0810.3472}}].

\bibitem{Murthy:2009dq}
S.~Murthy and B.~Pioline, {\it {A Farey tale for N=4 dyons}},  {\em JHEP} {\bf
  09} (2009) 022, [\href{http://arxiv.org/abs/0904.4253}{{\tt
  arXiv:0904.4253}}].

\bibitem{Dabholkar:2014ema}
A.~Dabholkar, J.~Gomes, and S.~Murthy, {\it {Nonperturbative black hole entropy
  and Kloosterman sums}},  {\em JHEP} {\bf 03} (2015) 074,
  [\href{http://arxiv.org/abs/1404.0033}{{\tt arXiv:1404.0033}}].

\bibitem{Maldacena:1999bp}
J.~M. Maldacena, G.~W. Moore, and A.~Strominger, {\it {Counting BPS black holes
  in toroidal Type II string theory}},
  [\href{http://arxiv.org/abs/hep-th/9903163}{{\tt hep-th/9903163}}].

\bibitem{Denef:2000nb}
F.~Denef, {\it {Supergravity flows and D-brane stability}},  {\em JHEP} {\bf
  08} (2000) 050, [\href{http://arxiv.org/abs/hep-th/0005049}{{\tt
  hep-th/0005049}}].

\bibitem{Dabholkar:2009dq}
A.~Dabholkar, M.~Guica, S.~Murthy, and S.~Nampuri, {\it {No entropy enigmas for
  N=4 dyons}},  {\em JHEP} {\bf 06} (2010) 007,
  [\href{http://arxiv.org/abs/0903.2481}{{\tt arXiv:0903.2481}}].

\bibitem{Maldacena:1996gb}
J.~M. Maldacena and A.~Strominger, {\it {Statistical entropy of
  four-dimensional extremal black holes}},  {\em Phys. Rev. Lett.} {\bf 77}
  (1996) 428--429, [\href{http://arxiv.org/abs/hep-th/9603060}{{\tt
  hep-th/9603060}}].

\bibitem{Dijkgraaf:2000fq}
R.~Dijkgraaf, J.~M. Maldacena, G.~W. Moore, and E.~P. Verlinde, {\it {A Black
  hole Farey tail}},  [\href{http://arxiv.org/abs/hep-th/0005003}{{\tt
  hep-th/0005003}}].

\bibitem{Gaiotto:2006wm}
D.~Gaiotto, A.~Strominger, and X.~Yin, {\it {The M5-Brane Elliptic Genus:
  Modularity and BPS States}},  {\em JHEP} {\bf 08} (2007) 070,
  [\href{http://arxiv.org/abs/hep-th/0607010}{{\tt hep-th/0607010}}].

\bibitem{deBoer:2006vg}
J.~de~Boer, M.~C.~N. Cheng, R.~Dijkgraaf, J.~Manschot, and E.~Verlinde, {\it {A
  Farey Tail for Attractor Black Holes}},  {\em JHEP} {\bf 11} (2006) 024,
  [\href{http://arxiv.org/abs/hep-th/0608059}{{\tt hep-th/0608059}}].

\bibitem{Manschot:2007ha}
J.~Manschot and G.~W. Moore, {\it {A Modern Farey Tail}},  {\em Commun. Num.
  Theor. Phys.} {\bf 4} (2010) 103--159,
  [\href{http://arxiv.org/abs/0712.0573}{{\tt arXiv:0712.0573}}].

\bibitem{Manschot:2009ia}
J.~Manschot, {\it {Stability and duality in N=2 supergravity}},  {\em Commun.
  Math. Phys.} {\bf 299} (2010) 651--676,
  [\href{http://arxiv.org/abs/0906.1767}{{\tt arXiv:0906.1767}}].

\bibitem{Alexandrov:2012au}
S.~Alexandrov, J.~Manschot, and B.~Pioline, {\it {D3-instantons, Mock Theta
  Series and Twistors}},  {\em JHEP} {\bf 04} (2013) 002,
  [\href{http://arxiv.org/abs/1207.1109}{{\tt arXiv:1207.1109}}].

\bibitem{Sen:2007qy}
A.~Sen, {\it {Black Hole Entropy Function, Attractors and Precision Counting of
  Microstates}},  {\em Gen. Rel. Grav.} {\bf 40} (2008) 2249--2431,
  [\href{http://arxiv.org/abs/0708.1270}{{\tt arXiv:0708.1270}}].

\bibitem{Dabholkar:2012nd}
A.~Dabholkar, S.~Murthy, and D.~Zagier, {\it {Quantum Black Holes, Wall
  Crossing, and Mock Modular Forms}},
  [\href{http://arxiv.org/abs/1208.4074}{{\tt arXiv:1208.4074}}].

\bibitem{Zwegers:2008zna}
S.~Zwegers, {\em {Mock Theta Functions}}.
\newblock PhD thesis, Utrecht University, 2008.
\newblock [\href{http://arxiv.org/abs/0807.4834}{{\tt arXiv:0807.4834}}].

\bibitem{Zagier:2007}
D.~Zagier, {\it {Ramanujan's mock theta functions and their applications
  [d'apr$\grave{\textrm{e}}$s Zwegers and Bringmann-Ono]}},  {\em
  {S{\'e}minaire BOURBAKI, 60 $\grave{e}me$ ann{\'e}e, 2006-2007}} {\bf {986}}
  (2007).

\bibitem{Bringmann:2010sd}
K.~Bringmann and J.~Manschot, {\it {From sheaves on $P^2$ to a generalization
  of the Rademacher expansion}}, [\href{http://arxiv.org/abs/1006.0915}{{\tt
  arXiv:1006.0915}}].

\bibitem{Huang:2015sta}
M.-x. Huang, S.~Katz, and A.~Klemm, {\it {Topological String on elliptic CY
  3-folds and the ring of Jacobi forms}},  {\em JHEP} {\bf 10} (2015) 125,
  [\href{http://arxiv.org/abs/1501.04891}{{\tt arXiv:1501.04891}}].

\bibitem{Haghighat:2015ega}
B.~Haghighat, S.~Murthy, C.~Vafa, and S.~Vandoren, {\it {F-Theory, Spinning
  Black Holes and Multi-string Branches}},
  [\href{http://arxiv.org/abs/1509.00455}{{\tt arXiv:1509.00455}}].

\bibitem{Gomes:2015xcf}
J.~Gomes, {\it {Exact holography and black hole entropy in N=8 and N=4 string
  theory}}, [\href{http://arxiv.org/abs/1511.07061}{{\tt arXiv:1511.07061}}].

\bibitem{Sen:2009vz}
A.~Sen, {\it {Arithmetic of Quantum Entropy Function}},  {\em JHEP} {\bf 08}
  (2009) 068, [\href{http://arxiv.org/abs/0903.1477}{{\tt arXiv:0903.1477}}].

\bibitem{Dabholkar:2010rm}
A.~Dabholkar, J.~Gomes, S.~Murthy, and A.~Sen, {\it {Supersymmetric Index from
  Black Hole Entropy}},  {\em JHEP} {\bf 04} (2011) 034,
  [\href{http://arxiv.org/abs/1009.3226}{{\tt arXiv:1009.3226}}].

\bibitem{Eichler:1985ja}
M.~Eichler and D.~Zagier, {\em {The Theory of Jacobi Forms}}.
\newblock Birkh{\"a}user, 1985.

\bibitem{Banerjee:2008ri}
S.~Banerjee and A.~Sen, {\it {S-duality Action on Discrete T-duality
  Invariants}},  {\em JHEP} {\bf 04} (2008) 012,
  [\href{http://arxiv.org/abs/0801.0149}{{\tt arXiv:0801.0149}}].

\bibitem{Dijkgraaf:1996it}
R.~Dijkgraaf, E.~P. Verlinde, and H.~L. Verlinde, {\it {Counting dyons in N=4
  string theory}},  {\em Nucl. Phys.} {\bf B484} (1997) 543--561,
  [\href{http://arxiv.org/abs/hep-th/9607026}{{\tt hep-th/9607026}}].

\bibitem{Shih:2005uc}
D.~Shih, A.~Strominger, and X.~Yin, {\it {Recounting Dyons in N=4 string
  theory}},  {\em JHEP} {\bf 10} (2006) 087,
  [\href{http://arxiv.org/abs/hep-th/0505094}{{\tt hep-th/0505094}}].

\bibitem{David:2006yn}
J.~R. David and A.~Sen, {\it {CHL Dyons and Statistical Entropy Function from
  D1-D5 System}},  {\em JHEP} {\bf 11} (2006) 072,
  [\href{http://arxiv.org/abs/hep-th/0605210}{{\tt hep-th/0605210}}].

\bibitem{Banerjee:2008pu}
S.~Banerjee, A.~Sen, and Y.~K. Srivastava, {\it {Partition Functions of Torsion
  > 1 Dyons in Heterotic String Theory on $T^6$}},  {\em JHEP} {\bf 05} (2008)
  098, [\href{http://arxiv.org/abs/0802.1556}{{\tt arXiv:0802.1556}}].

\bibitem{Dabholkar:2008zy}
A.~Dabholkar, J.~Gomes, and S.~Murthy, {\it {Counting all dyons in N =4 string
  theory}},  {\em JHEP} {\bf 05} (2011) 059,
  [\href{http://arxiv.org/abs/0803.2692}{{\tt arXiv:0803.2692}}].

\bibitem{Cheng:2007ch}
M.~C.~N. Cheng and E.~Verlinde, {\it {Dying Dyons Don't Count}},  {\em JHEP}
  {\bf 09} (2007) 070, [\href{http://arxiv.org/abs/0706.2363}{{\tt
  arXiv:0706.2363}}].

\bibitem{Dabholkar:2007vk}
A.~Dabholkar, D.~Gaiotto, and S.~Nampuri, {\it {Comments on the spectrum of CHL
  dyons}},  {\em JHEP} {\bf 01} (2008) 023,
  [\href{http://arxiv.org/abs/hep-th/0702150}{{\tt hep-th/0702150}}].

\bibitem{Troost:2010ud}
J.~Troost, {\it {The non-compact elliptic genus: mock or modular}},  {\em JHEP}
  {\bf 06} (2010) 104, [\href{http://arxiv.org/abs/1004.3649}{{\tt
  arXiv:1004.3649}}].

\bibitem{Eguchi:2010cb}
T.~Eguchi and Y.~Sugawara, {\it {Non-holomorphic Modular Forms and SL(2,R)/U(1)
  Superconformal Field Theory}},  {\em JHEP} {\bf 03} (2011) 107,
  [\href{http://arxiv.org/abs/1012.5721}{{\tt arXiv:1012.5721}}].

\bibitem{Ashok:2011cy}
S.~K. Ashok and J.~Troost, {\it {A Twisted Non-compact Elliptic Genus}},  {\em
  JHEP} {\bf 03} (2011) 067, [\href{http://arxiv.org/abs/1101.1059}{{\tt
  arXiv:1101.1059}}].

\bibitem{Murthy:2013mya}
S.~Murthy, {\it {A holomorphic anomaly in the elliptic genus}},  {\em JHEP}
  {\bf 06} (2014) 165, [\href{http://arxiv.org/abs/1311.0918}{{\tt
  arXiv:1311.0918}}].

\bibitem{Ashok:2013pya}
S.~K. Ashok, N.~Doroud, and J.~Troost, {\it {Localization and real Jacobi
  forms}},  {\em JHEP} {\bf 04} (2014) 119,
  [\href{http://arxiv.org/abs/1311.1110}{{\tt arXiv:1311.1110}}].

\bibitem{Harvey:2014nha}
J.~A. Harvey, S.~Lee, and S.~Murthy, {\it {Elliptic genera of ALE and ALF
  manifolds from gauged linear sigma models}},  {\em JHEP} {\bf 02} (2015) 110,
  [\href{http://arxiv.org/abs/1406.6342}{{\tt arXiv:1406.6342}}].

\bibitem{Pioline:2015wza}
B.~Pioline, {\it {Wall-crossing made smooth}},  {\em JHEP} {\bf 04} (2015) 092,
  [\href{http://arxiv.org/abs/1501.01643}{{\tt arXiv:1501.01643}}].

\bibitem{Bringmann:2012zr}
K.~Bringmann and S.~Murthy, {\it {On the positivity of black hole degeneracies
  in string theory}},  {\em Commun. Num. Theor Phys.} {\bf 07} (2013) 15--56,
  [\href{http://arxiv.org/abs/1208.3476}{{\tt arXiv:1208.3476}}].

\bibitem{Gupta:2012cy}
R.~K. Gupta and S.~Murthy, {\it {All solutions of the localization equations
  for N=2 quantum black hole entropy}},  {\em JHEP} {\bf 02} (2013) 141,
  [\href{http://arxiv.org/abs/1208.6221}{{\tt arXiv:1208.6221}}].

\bibitem{Murthy:2013xpa}
S.~Murthy and V.~Reys, {\it {Quantum black hole entropy and the holomorphic
  prepotential of N=2 supergravity}},  {\em JHEP} {\bf 10} (2013) 099,
  [\href{http://arxiv.org/abs/1306.3796}{{\tt arXiv:1306.3796}}].

\bibitem{Gupta:2015gga}
R.~K. Gupta, Y.~Ito, and I.~Jeon, {\it {Supersymmetric Localization for BPS
  Black Hole Entropy: 1-loop Partition Function from Vector Multiplets}},
  [\href{http://arxiv.org/abs/1504.01700}{{\tt arXiv:1504.01700}}].

\bibitem{Sen:2011ba}
A.~Sen, {\it {Logarithmic Corrections to N=2 Black Hole Entropy: An Infrared
  Window into the Microstates}},  {\em Gen. Rel. Grav.} {\bf 44} (2012), no.~5
  1207--1266, [\href{http://arxiv.org/abs/1108.3842}{{\tt arXiv:1108.3842}}].

\bibitem{Cardoso:2008fr}
G.~L. Cardoso, B.~de~Wit, and S.~Mahapatra, {\it {Subleading and
  non-holomorphic corrections to N=2 BPS black hole entropy}},  {\em JHEP} {\bf
  02} (2009) 006, [\href{http://arxiv.org/abs/0808.2627}{{\tt
  arXiv:0808.2627}}].

\bibitem{LopesCardoso:2006bg}
G.~Lopes~Cardoso, B.~de~Wit, J.~Kappeli, and T.~Mohaupt, {\it {Black hole
  partition functions and duality}},  {\em JHEP} {\bf 03} (2006) 074,
  [\href{http://arxiv.org/abs/hep-th/0601108}{{\tt hep-th/0601108}}].

\bibitem{LopesCardoso:2004xf}
G.~Lopes~Cardoso, B.~de~Wit, J.~Kappeli, and T.~Mohaupt, {\it {Asymptotic
  degeneracy of dyonic N = 4 string states and black hole entropy}},  {\em
  JHEP} {\bf 12} (2004) 075, [\href{http://arxiv.org/abs/hep-th/0412287}{{\tt
  hep-th/0412287}}].

\end{thebibliography}
\end{document}